\documentstyle[aps,epsfig,rotate]{revtex}
\begin{document}
\newcommand{\pom}{{I\!\!P}}
\newcommand{\Pom}{{I\!\!P}}
\newcommand{\regg}{{I\!\!R}}
\newcommand{\Regg}{{I\!\!R}}
\def\lsim{\mathrel{\rlap{\lower4pt\hbox{\hskip1pt$\sim$}}
    \raise1pt\hbox{$<$}}}                % less than or approx. symbol
\def\gsim{\mathrel{\rlap{\lower4pt\hbox{\hskip1pt$\sim$}}
    \raise1pt\hbox{$>$}}}                % greater than or approx. symbol
\newcommand{\qsq} {$Q^2$}
\newcommand{\wsq} {$W^2$}
\newcommand{\xpom} {$x_\Pom$}
\newcommand{\x} {$x$}
\newcommand{\xbj} {$x_{Bj}$}
\newcommand{\xl} {$x_L$}
\newcommand{\W} {$W$}
\newcommand{\ttt} {$t$}
\newcommand{\tmod} {| t | } 
\newcommand{\etmax} {$\eta_{max}$}
\newcommand{\bet} {$\beta$}
\newcommand{\X} {$X$}
\newcommand{\mx} {$M_X$}
\newcommand{\Y} {$Y$}
\newcommand{\my} {$M_Y$}
\newcommand{\mpro} {$M_p$}
\newcommand{\pbinv} {$\rm pb^{-1}$}
\newcommand{\nbinv} {$\rm nb^{-1}$}
\newcommand{\gev} {\rm GeV}
\newcommand{\Gev} {\rm GeV}
\newcommand{\gevsq} {$\rm GeV^2$}
\newcommand{\Gevsq} {$\rm GeV^2$}
\newcommand{\gevsqm} {$\rm GeV^{-2}$}
\newcommand{\Gevsqm} {$\rm GeV^{-2}$}
\newcommand{\jpsi} {$J / \psi$}
%!!!!!!!!!!!!!!!!!!!!!!! box pour F2, A, etc
\newcommand{\fdthree} {$F_2^{D(3)}$}
\newcommand{\fdthreef} {$F_2^{D(3)}  (Q^2, x_{I\!\!P} , \beta )$}
\newcommand{\fdfour} {$F_2^{D(4)}$}
\newcommand{\fdfourf} {$F_2^{D(4)}  (Q^2, x_{I\!\!P} , \beta , t)$}
\newcommand{\fdc} {F_2^{D}_{charm}}
\newcommand{\flpom} {F_L^{\Pom}}
\newcommand{\RD} {$R_D$}
\newcommand{\A} {$A (Q^2, \beta)$}
\newcommand{\Apom} {$A_\pom (Q^2, \beta)$}
\newcommand{\Aregg} {$A_\regg (Q^2, \beta)}

%                                                    journals
\def\ar#1#2#3   {{\em Ann. Rev. Nucl. Part. Sci.} {\bf#1} (#2) #3}
\def\epj#1#2#3  {{\em Eur. Phys. J.} {\bf#1} (#2) #3}
\def\err#1#2#3  {{\it Erratum} {\bf#1} (#2) #3}
\def\ib#1#2#3   {{\it ibid.} {\bf#1} (#2) #3}
\def\ijmp#1#2#3 {{\em Int. J. Mod. Phys.} {\bf#1} (#2) #3}
\def\jetp#1#2#3 {{\em JETP Lett.} {\bf#1} (#2) #3}
\def\mpl#1#2#3  {{\em Mod. Phys. Lett.} {\bf#1} (#2) #3}
\def\nim#1#2#3  {{\em Nucl. Instr. Meth.} {\bf#1} (#2) #3}
\def\nc#1#2#3   {{\em Nuovo Cim.} {\bf#1} (#2) #3}
\def\np#1#2#3   {{\em Nucl. Phys.} {\bf#1} (#2) #3}
\def\pl#1#2#3   {{\em Phys. Lett.} {\bf#1} (#2) #3}
\def\prep#1#2#3 {{\em Phys. Rep.} {\bf#1} (#2) #3}
\def\prev#1#2#3 {{\em Phys. Rev.} {\bf#1} (#2) #3}
\def\prl#1#2#3  {{\em Phys. Rev. Lett.} {\bf#1} (#2) #3}
\def\ptp#1#2#3  {{\em Prog. Th. Phys.} {\bf#1} (#2) #3}
\def\rmp#1#2#3  {{\em Rev. Mod. Phys.} {\bf#1} (#2) #3}
\def\rpp#1#2#3  {{\em Rep. Prog. Phys.} {\bf#1} (#2) #3}
\def\sjnp#1#2#3 {{\em Sov. J. Nucl. Phys.} {\bf#1} (#2) #3}
\def\spj#1#2#3  {{\em Sov. Phys. JEPT} {\bf#1} (#2) #3}
\def\zp#1#2#3   {{\em Zeit. Phys.} {\bf#1} (#2) #3}
%

%%%%%%%%%%%%%%%%%%%%%%%%%%%%%%%%%%%%%%%%
%%%%%%%%%%%%%%%%%%%%%%%%%%%%%

\title {Diffraction at HERA \\
An Experimentalist's View}

\author{
    P. Marage}
\address{
    Universit\'e Libre de Bruxelles $-$ CP 230 \\
   Boulevard du Triomphe  \\
    B-1050 Brussels Belgium}
    
\date{\today}

\maketitle{}

\begin{abstract}

Diffraction studies at HERA are introduced, with reference to other 
communications to this Conference. 
Motivations and specific features of the experimental
approaches are stressed.\footnote
{Paper presented at the LAFEX International School on High Energy 
Physics, LISHEP 98, Rio de Janeiro, February 1998. 
The present paper follows closely the content of the talk presented at Rio, 
with the exception that reference to preliminary results have been updated 
to published papers where available; a few additions have
been included, mainly in the form of footnotes.}
 
\end{abstract}

%%%%%%%%%%%%%%%%%%%%%%%%%%%%%%%%%%%%
%%%%%%%%%%%%%%%%%%%%%%%

%====================================================
%====================================================
\section {Introduction}
      							\label{sect:intro}

\begin{figure}[tbp]
\vspace{-0.5cm}
\begin{center}
\epsfig{file=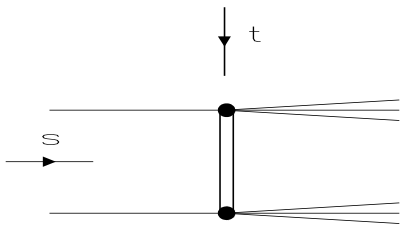,width=8cm,height=4cm}
\end{center}
\vspace{-0.5cm}
\caption{Diffractive scattering of hadrons, with double diffractive 
dissociation.}
\label{fig:diffr}
\end{figure}

Diffractive scattering of hadrons (see Fig. \ref{fig:diffr}) is closely 
related to elastic scattering and thus to deepest 
principles of quantum theory, wave-particle duality and unitarity, 
through the optical theorem and the Froissard and Pumplin bounds (see e.g. 
\cite{Collins,Kaidalov_phys_rep}).
%In spite of this fundamental importance, theoretical difficulties and the 
%contrasting succes of the parton approach led to a lack of interest for 
%diffraction studies in the end 70's and the 80's, until the HERA and 
%Tevatron revival in the last years \cite {Predazzi}.

%=======================\label{fig:diffr} ====================
%=======================\end{fig:diffr} =======================

In the $s$-channel approach, diffractive scattering is explained by the 
differential absorption by the target of the large number of hadronic states 
which coherently build up the initial state hadron and scatter with different 
cross sections \cite{GoodWalker}.
The implied reorganisation in the outgoing system leads to the production 
of hadron states with different mass and / or quantum numbers, but obeying 
definite selection rules.
Whereas elastic scattering is intimately related to the amount of inelastic 
scattering (optical theorem) and happens over the full target volume, diffractive 
scattering is due to fluctuations in the inelastic amplitudes.
It is of a peripheral character, since the absorption cross sections for 
the different hadronic states vary more in the outer (``grey'') region of 
the target than in the center (``black'') core.
%, where they are fully absorbed.
%When the $t$ distribution of the cross section is parameterised as an 
%exponential of the form $e^{b \cdot t}$ ($t$ is the squared four-momentum 
%of the exchange), the parameter $b$ is thus larger in the case of diffractive 
%dissociation than for elastic scattering.
%However, 
A limitation of the $s$-channel approach is that it does not provide {\it ab 
initio} calculations but requires, for any practical purpose, the use of badly 
known cross sections and of non-diagonal scattering properties.

In the $t$-channel approach \cite{Collins}, the scattering properties are
related to the characteristics of exchanged virtual particles
(with $t$ the squared four-momentum of the exchange), using general
properties of unitarity, crossing symmetry and analyticity of the amplitudes.
In particular, the energy dependence of the cross section is governed by the
spin of the exchanged particles.
In a generalised form, this leads to the concept of ``Regge trajectories'': in 
the $t$ $-$ angular momentum plane, real particles (with squared mass $t > 0$) 
and virtual states (with $t < 0$) are observed to lie on linear trajectories, 
characterised by a definite set of quantum numbers and parameterised as 
\begin{equation}
\alpha(t) = \alpha(0) + \alpha ^{\prime} \cdot t  .
							\label{eq:traj} 
\end{equation}
For the exchange of a given trajectory, the cross section depends on the 
centre of mass energy squared $s$ as
\begin{equation}
%\frac {{\rm d} \sigma}{{\rm d} t}  \propto s^{2 \cdot \alpha(t) - 1} 
         \sigma \propto s^{\alpha(0) - 1}.
							\label{eq:en_dep}  
\end{equation}

This approach met tremendous success in the description of total, elastic, 
single- and double-diffraction scattering cross sections (see e.g. 
\cite{Kaidalov_phys_rep,alberi_phys_rep,Goulianos_phys_rep,Predazzi}).
In particular, the celebrated Donnachie-Landshoff parameterisation 
\cite{DoLa} describes the elastic cross section for numerous 
reactions in terms of the exchange of two main trajectories: the {\it 
reggeon} trajectory, related to the $\rho$ meson family, of the form
\begin{equation}
 \alpha _\regg (t) \simeq\ 0.55 + 0.9 \ t ,
							\label{eq:traj_regg} 
\end{equation}
$t$ being measured in \gevsq, and the {\it pomeron} trajectory, of the form
\footnote{An analysis by Cudell et al. \cite{Cudell} gives for the 
pomeron intercept a preferred value $\alpha_\pom(0) \simeq\ 1.10$, with extreme
acceptable values of 1.07 and 1.11.}
\begin{equation}
 \alpha _\pom (t) \simeq\ 1.08 + 0.25 \ t   ,
							\label{eq:traj_pom} 
\end{equation}
which carries the quantum numbers of the vacuum and to which no known 
particle is associated (except maybe for a glueball candidate \cite{WA91}).
At high energy, elastic and diffractive scattering are dominated by pomeron 
exchange, which leads to a slow increase of the cross section with $s$.

In spite of this success, the need of considering multi-reggeon and cut 
exchange led to intricated mathematics.
More fundamentally, questions concerning the nature of the pomeron and 
the lack of a microscopic theory led in the 70's to {\it ``the death of the 
Reggeon approach''} \cite{Levin_the_death}.

With the advent of QCD as the theory of strong interactions, models were 
proposed in order to understand diffraction in this framework. 
In the simplest form, the pomeron was modelised as a two-gluon system 
\cite{low_nussinov}, and this approach was actively pursued by several authors.
%\cite{diffr_in_QCD}.
At the end of the 80's, the 
observation by the UA8 experiment of jet production in diffractive $p \bar 
p$ scattering \cite{UA8} confirmed the assumption by Ingelman and Schlein 
\cite{Ingelman-Schlein} that the pomeron may be built of partons subjected to 
hard scattering.

The major discovery of the ZEUS and H1 experiments in deep inelastic 
scattering (DIS) was the observation in 1992 that the proton structure function 
$F_2 (Q^2, x)$ is sharply rising for small $x$ values (i.e. high energy).
This ``hard'' behaviour, observed even for low \qsq\ values 
\cite{H1_f2,ZEUS_f2} contrasts with the ``soft'' behaviour of high-energy
hadron$-$hadron cross sections (eq. \ref{eq:traj_pom}).

%=======================\label{fig:diffr_event} =================
\begin{figure}[tbp]
\vspace*{1.cm}
\begin{center}
\epsfig{file=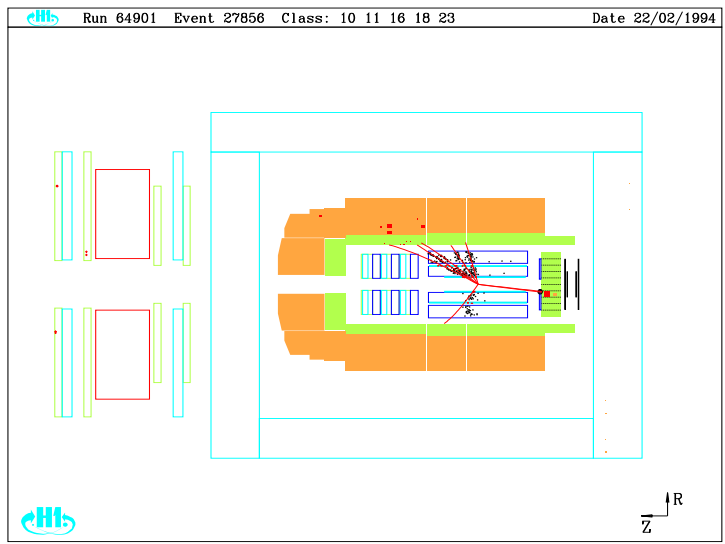,width=5cm,height=8cm,angle=90}
\end{center}
\vspace{1.5cm}
\caption{A typical diffractive DIS event in the H1 detector. 
The electron is scattered on the right, in the backward 
electromagnetic calorimeter.
No activity is detected in the forward part of the liquid argon  
calorimeter nor in the forward detectors, in particular the forward
muon detectors, on the left.}
\label{fig:diffr_event}
\end{figure}
%=======================\end{fig:diffr_event} ===================

A second major result was the
{\it ``Observation of Events with a Large Rapidity Gap in Deep Inelastic 
Scattering at HERA''} by ZEUS using a sample of $\simeq$ 25 \nbinv\ 
accumulated in 1992 \cite{ZEUS_diffr_1992}, confirmed by H1 with
300 \nbinv\ of data taken in 1993 \cite{H1_diffr_1993}.
These large rapidity gap (LRG) events, attributed to diffraction, were observed 
to make a contribution to the DIS cross section at a level of $8 - 10 \%$, 
with a leading twist \qsq\ dependence.
They are characterised (see an example in Fig. \ref{fig:diffr_event}) by 
the absence of activity in the ``forward'' part of the detectors.\footnote
{In the HERA convention, the ``forward'' ($+z$) direction 
is that of the outgoing proton beam; unless stated otherwise, the transverse 
direction is defined here with respect to the beam direction.}
In contrast, for non-diffractive ``usual'' DIS events, a colour string 
extends through the forward region, connecting the proton remnant and the struck 
quark, which leads to particle emission.

It is true to say\cite{Kowalski_Rome} that this observation was a surprise 
for many experimentalists: few papers dealing with diffraction in DIS were 
presented at the 1987 and 1991 Workshops on HERA physics 
\cite{Workshops}, no diffractive Monte Carlo in DIS was available, and the 
detectors were not well equipped for diffractive studies (nor for low-\x\ 
physics in general).
It was also true to state, as written in the first experimental HERA paper 
on diffraction, that ``until recently, Regge theory and perturbative QCD 
have been subjects without much overlap'' \cite{ZEUS_diffr_1992}.

In the last few years, however, considerable progress have been made in both  
experimental and theoretical research, as testified by the large number of 
published papers and of workshops devoted to diffraction.
In addition to HERA, diffraction has been intensively studied at the
Fermilab Tevatron \cite{tevatron}. 
Interactions between experimentalists and theorists have intensively 
developed, the complexity of the subject making the guidance of experiment 
important for the progress of theory.

%====================================================
%====================================================
\section {Inclusive Measurements of the Diffractive Cross Section}
      							\label{sect:incl}

Huge efforts have been invested by the experimental collaborations to achieve 
a precise measurement of the inclusive diffraction cross section at HERA.
%, which makes a fundamental contribution to the bulk of $e p$ interactions.
Common definitions of the relevant variables have been 
accepted, the concept of ``diffractive structure functions'' has emerged as 
a useful tool and, most important, experimental procedures and their 
inter-relations have been discussed and clarified (in particular the 
question of non-diffractive background subtractions), in order to define 
precisely and unambiguously the object of the studies.
%The measurement of the differential diffractive cross sections, together
%with the study of diffractive final states and of semi-inclusive diffractive 
%processes, may provide an access to the understanding of the partonic content 
%of the pomeron and thus to a ``microscopic'' picture of diffraction in terms 
%of QCD.

%====================================================
%====================================================
\subsection {Kinematics; Diffractive Structure Functions}
      							\label{sect:kin}

%=======================\label{fig:rapidity_gap}===============
\begin{figure}[tbp]
\vspace{-0.5cm}
\begin{center}
\epsfig{file=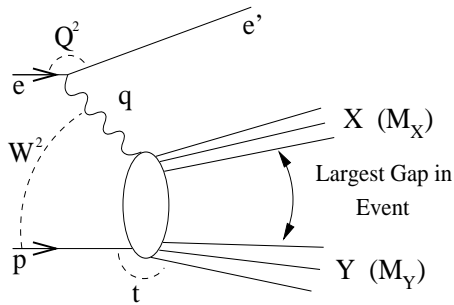,width=4cm,height=6cm,angle=270}
\end{center}
\vspace{0.cm}
\caption{Deep-inelastic diffractive interaction.}
\label{fig:rapidity_gap}
\end{figure}
%=======================\end{fig:rapidity_gap} ======================

Diffractive $e p$ interactions are sketched in Fig. \ref{fig:rapidity_gap}.
The characteristic feature is that the final state hadronic system, with 
centre of mass energy \W,\footnote
{Here, the mass \W\ is always supposed 
{\it large}, i.e. $W \gg M_p $, \mpro \ being the proton mass. Note that the
centre of mass hadronic energy is called $W$ in deep-inelastic scattering
and $\sqrt{s}$ for hadron$-$hadron interactions; in DIS, $\sqrt{s}$ is the 
$e - p$ centre of mass energy.}
is divided into two subsystems of significantly lower mass, separated by a 
large gap in rapidity: the system $X$, of mass \mx, which corresponds to photon 
dissociation, and the system $Y$, of mass \my, which consists in a proton or 
an excited baryonic state, with small transverse momentum 
$p_T \simeq\ \sqrt{|t|}$.
An important feature is that the size of the gap in rapidity is significantly
larger than implied by particle density fluctuation during the 
hadronisation process for non-diffractive interactions, and
is thus attributed to the exchange of a colour-singlet system, specifically 
reggeon or pomeron.\footnote
{As will be mentioned below (section \ref{sect:SCI}), approaches have also
been proposed for LRG event production which do not refer to
diffraction as a specific process, but are based on colour reorganisation
through soft processes.}

In the particular case where the proton remains intact, the diffractive 
process
\begin{equation}
e + p \rightarrow e + X + p
						\label{eq:ep_reaction}
\end{equation}
is defined, up to an azimuthal angle between the electron and the proton scattering 
planes, by four kinematical variables.
These are conveniently chosen as \qsq , \xpom , $\beta$ and \ttt , where \qsq\ is the 
negative of the squared four-momentum of the photon, \ttt \ is the squared 
four-momentum transfer to the proton, and \xpom \ and $\beta$ are defined as
%
%\begin{eqnarray}
\begin{equation}
   x_\pom = 1 - x_L \simeq \frac{Q^2 + M_X^2}{Q^2 + W^2 }
						\label{eq:xpom_def}
\end{equation}

\begin{equation}
    \beta \simeq \frac{Q^2 }{Q^2 + M_X^2} ,
						\label{eq:beta_def}
\end{equation}
%\end{eqnarray}
%
\xl \ being the fraction of the incident proton energy carried by the 
scattered proton. 

The variable \xpom\ can be interpreted, in the proton infinite momentum 
frame, as the fraction of the proton momentum carried by the exchange (reggeon 
or pomeron) and $\beta $ is the fraction of the exchanged momentum carried by 
the quark struck by the photon.
These variables are related to the $x$ scaling variable (with 
$W^2 \simeq Q^2 / x - Q^2$) by the relation
\begin{equation}
x = \beta \cdot x_\pom  . 
						\label{eq:x} 
\end{equation}
Kinematics imply that, at high energy, a large gap in rapidity is created between
the system $X$ and the scattered proton when $x_\pom \ll 1$, i.e. $M_X \ll W$.

In analogy with non-diffractive DIS scattering, the measured cross section 
is expressed in the form of a four-fold {\it diffractive structure function} 
\fdfourf :
% \cite{H1_f2d_93,Kowalski_talk} : 
%
\begin{equation}
  \frac { {\rm d}^4 \sigma \ (e + p \rightarrow e + X + p) } 
 { {\rm d}Q^2 \ {\rm d}x_{I\!\!P} \ {\rm d}\beta \ {\rm d}t } 
        = \frac {4 \pi \alpha^2} {\beta Q^4} 
            \ (1 - y + \frac {y^2} {2 (1 + R_D) } )
            \ F_2^{D(4)} (Q^2, x_{I\!\!P} , \beta , t) ,
                                            \label{eq:fdfourfull}
\end{equation}
where $y$ is the usual scaling variable, with
$  y \simeq W^2 / s, $
and \RD\ is the ratio of the longitudinal and transverse diffractive cross 
sections. 
\RD\ has not been measured so far.\footnote
{The measurement of $R_D$ would be of great interest, since various models make
different predictions for the longitudinal cross section (see the review
\cite{Briskin}).}
Its value is commonly put to 0 for extracting the diffractive structure function,
which has a small impact on the measurements performed in the presently 
accessible $y$ range.

Experimentally, the $t$ variable is usually not measured or is integrated over. 
Results are thus mostly presented for the three-fold diffractive structure 
function \fdthreef .
The latter is conveniently parameterised in the {\it factorised} form
\begin{equation}
F_2^{D(3)} (Q^2, x_{I\!\!P} , \beta ) = 
     \Phi (x_{I\!\!P} ) \cdot F_2^D (Q^2, \beta) , 
                                               \label{eq:factoris} 
\end{equation}
with a Regge inspired parameterisation:
\begin{equation}
  \Phi(x_\pom ) \propto x_\pom ^n
                                              \label{eq:flux_factor} 
\end{equation}
and
\begin{equation}
  n = 2 \cdot \langle \alpha (t) \rangle - 1.
                                              \label{eq:n} 
\end{equation}
For pomeron exchange, and if factorisation holds, the factor $\Phi(x_\pom )$
can be interpreted as describing an effective pomeron flux in the proton, whereas 
the function $F_2^D (Q^2, \beta)$ describes the pomeron structure,
$\beta$ playing the role of $x$ for hadron structure functions.

Several methods are used to extract the diffractive (i.e. pomeron 
exchange)\footnote 
{The term diffraction is often used, in a wide sense, for events with
a large rapidity gap; strictly speaking, it applies to pomeron exchange.} 
cross section from the deep-inelastic data.
The challenge for experimentalists is to measure a well-defined quantity,
not depending on models for background subtraction, and to 
minimise the contamination of non-diffractive events and of diffractive 
events with proton dissociation.

%====================================================
%====================================================
\subsection {Rapidity Gap Measurement Method; Reggeon Contribution}
      							\label{sect:rapgap}

%====================\label{fig:H1_f2d_94}==================
\begin{figure}[tbp]
\vspace{-1.cm}
\begin{center}
\epsfig{file=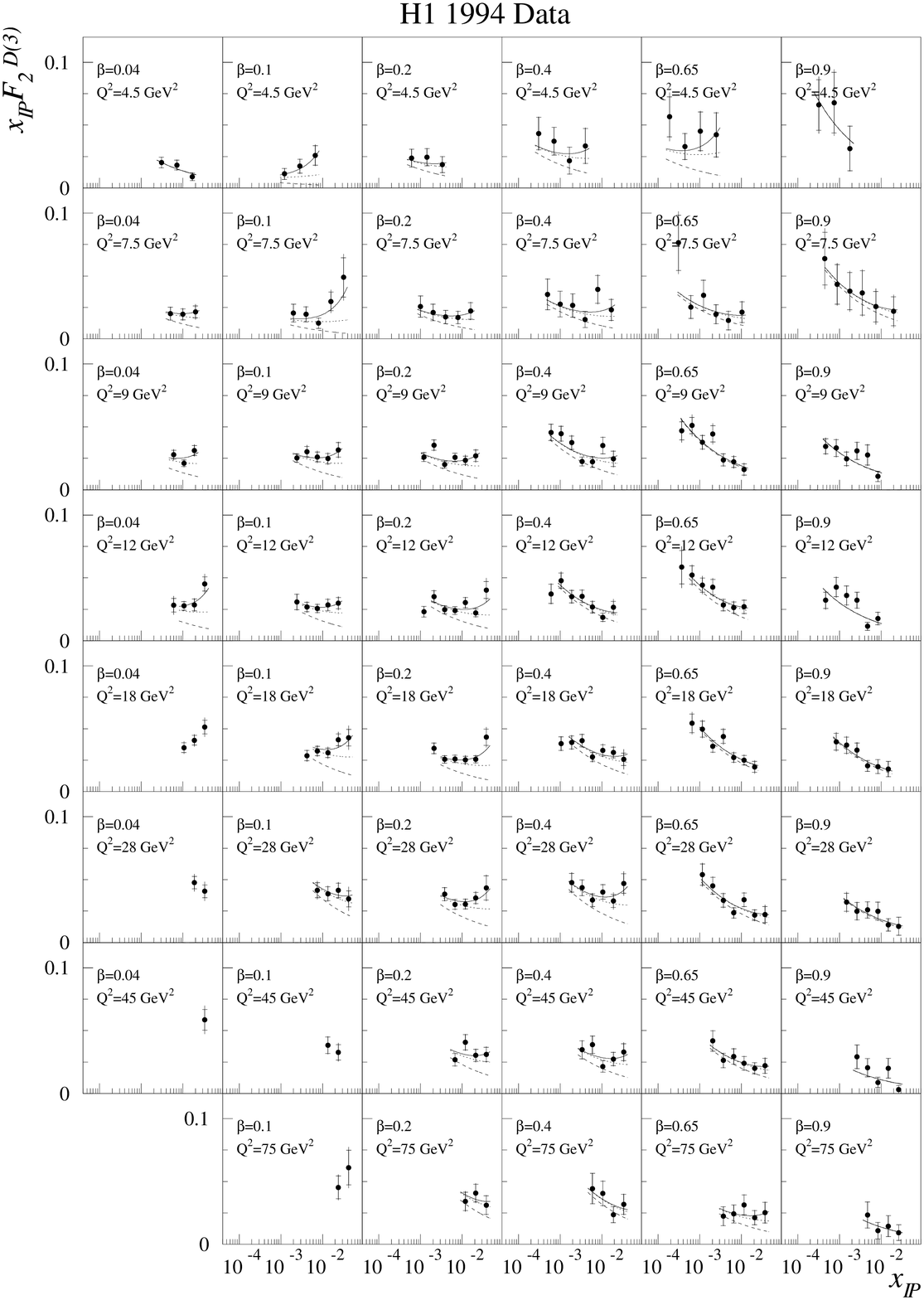,width=16cm,height=20cm}
\end{center}
\vspace{0.cm}
\caption{H1 measurement of $x_\pom\ \cdot$ \fdthreef\ ($M_Y < 1.6$ \gev ,
$|t| < 1 $ \gevsq ) as a function of $x_\pom$
for various \qsq\ and $\beta$ values, with the 1994 data.
The curves show the results of the Regge fit with interference. 
The dashed curves show the contributions from the pomeron alone, the dotted 
curves, the pomeron plus interference, and the continuous curves, the total.}
\label{fig:H1_f2d_94}
\end{figure}
%====================\end{fig:H1_f2d_94} ================

A natural way to study diffraction experimentally is to select events 
with a large rapidity gap, since the latter is kinematically related to small 
values of $x_\pom $ and thus, at high energy, to pomeron exchange.

This procedure is used by the H1 experiment, which takes advantage of a 
good coverture of the forward region.
Specifically, events are  selected, for which hadronic activity is observed 
in the central detectors, whereas the pseudorapidity \footnote 
{The pseudorapidity $\eta$ of a detected object is defined as 
$\eta = - \log \ \tan (\theta / 2) $,
$\theta$ being the emission angle defined with respect to the outgoing 
proton beam direction.}
of the most forward track or 
energy deposit in the central calorimeter is $ \eta_{max} = 3.2 $ 
and no activity is registered in the ``forward detectors'', i.e. the ``plug'' 
calorimeter, the forward muon detectors and the proton remnant tagger.
These detectors overlap in acceptance and are sensitive to primary 
particles emitted at small angle and to secondaries due to rescatterings 
in the beam pipe wall or adjacent collimators.
For these events, no hadron emission is thus observed in the 
large rapidity region $3.2 < \eta < 7.5$.
The selected sample is consequently
restricted to events with an elastically scattered proton or a low mass 
($M_Y \lsim 1.6 $ \gev) proton dissociation system (the latter contribute a few 
\% of the selected events), with $|t| \lsim 1$ \gevsq.

The ZEUS experiment has similarly selected, in the 1993 data sample, 
events with $ \eta _{max} = 2.5 $, using only the main calorimeter.
The implied relatively small gap in rapidity (the calorimeter extends up 
to $ \eta \simeq 3.8 $) required the use of Monte Carlo simulations to 
estimate the contamination of non-diffractive events with a gap 
due to particle density fluctuation during hadronisation and of events with 
proton dissociation.

With the integrated luminosity of $300 - 500$ \nbinv\ accumulated in 1993, 
the H1 and ZEUS experiments could present a first measurement of the 
diffractive structure function \fdthree \ as a function of \xpom, for different 
bins in \qsq \ and $\beta$ \cite{H1_f2d_93,ZEUS_f2d_93}. Within the 
measurement precision, factorisation in the sense of eqs. \ref{eq:factoris} 
and \ref{eq:flux_factor} was observed.

The integrated luminosity of 2 \pbinv\ accumulated in 1994 by the H1 
experiment allowed the measurement of \fdthree \ in a total of 47 bins in 
$\beta$ and \qsq \ ($0.04 < \beta < 0.9$ and $4.5 < Q^2 < 75 $ \gevsq ) with 
\xpom\ $ < 0.05$, the kinematically accessible \xpom \ range varying from 
bin to bin (see Fig. \ref{fig:H1_f2d_94}) \cite{H1_f2d_94}.
This measurement has been extended by H1 to low \qsq\ values 
($0.4 < Q^2 < 5$ \gevsq ) using the 1995 data with a vertex shifted 
towards the forward direction in the detector, and to $200 < Q^2 < 800$ \gevsq, 
using the statistics accumulated in $1995-1997$ \cite{H1_f2d_SV}.

With the increased precision of these measurements, presented in the sensitive 
form of the $x_\pom\ \cdot $ \fdthree \ distribution, the apparent factorisation
of eq. \ref{eq:factoris} was broken.
This feature was explained as due to the superposition of two contributions, 
corresponding respectively to pomeron and reggeon exchange:
\begin{eqnarray}
F_2^{D(3)} \ (Q^2, x_{I\!\!P} , \beta )
	& =  & \Phi^\pom (x_\pom) \cdot F_2^\pom (Q^2,\beta) + 
              \Phi^\regg (x_\pom) \cdot  F_2^\regg (Q^2,\beta) +
	      {\rm interf.}                                 
	                                              \nonumber \\
	& =  & x_\pom^{2 \cdot \langle \alpha_\pom \rangle - 1} 
                               \cdot F_2^\pom (Q^2,\beta)
	    + x_\pom^{2 \cdot \langle \alpha_\regg \rangle - 1} 
                               \cdot F_2^\regg (Q^2,\beta)
	    + {\rm interf.}
							\label{eq:factoris2}
\end{eqnarray}
%

%=======================\label{fig:two_bins} =================
\begin{figure}[tbp]
\vspace{-0.5cm}
\begin{center}
\epsfig{file=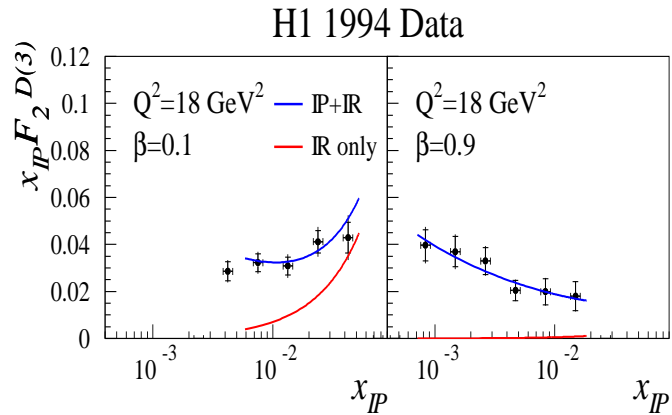,width=10cm,height=6.25cm}
\end{center}
\vspace{0. cm}
\caption{H1 measurement of $x_\pom\ \cdot$ \fdthreef\ as a function of $x_\pom$
for two \qsq\ and $\beta$ values, with the 1994 data. 
The lower curves correspond to the reggeon contribution only, the upper curves 
to the summed reggeon and pomeron contributions.}
\label{fig:two_bins}
\end{figure}
%=======================\end{fig:two_bins} ==================

From a Regge fit of the data to eq. \ref{eq:factoris2}, 
the reggeon trajectory intercept is found to be 
$\alpha_\regg (0) = 0.50 \pm 0.18$, in agreement with the expected value 
(eq. \ref{eq:traj_regg}), and the pomeron intercept is measured to be 
$\alpha_\pom (0) = 1.20 \pm 0.04$, higher than for soft interactions 
(see discussion in section \ref{sect:soft_hard}).
In the case of an incoming virtual state, the 
strength of the interference between pomeron and reggeon exchange is not 
known \cite{Marage_Madrid}; the data are compatible with maximum as well as 
with no interference \cite{H1_f2d_94}.

The detailed contributions of pomeron and reggeon exchange are illustrated
on Fig. \ref{fig:two_bins}.
The reggeon contribution is larger for larger values of 
\xpom, which correspond to smaller energy (for given \qsq\ and $\beta$
values, $x_\pom = x / \beta \simeq Q^2 / [\beta \cdot W^2] )$.
It is also larger for small values of $\beta$, which is consistent with
the expected decrease with $\beta$ of the reggeon structure function, following
the meson example, whereas the pomeron structure function is observed to 
be approximately flat in $\beta$ (see Fig. \ref{fig:scal_viol_beta} below).

Within the precision of the measurement, no evidence is found for 
factorisation breaking of the pomeron term itself, also
when the lower \qsq\ data are included \cite{H1_f2d_SV}.

%====================================================
%====================================================

%\subsection {The {\boldmath $\ln M_X^2 $} Measurement Method}
\subsection {The $\ln M_X^2 $ Measurement Method}
      							\label{sect:logmx}

\begin{figure}[tbp]
\vspace{0.cm}
\begin{center}
\epsfig{file=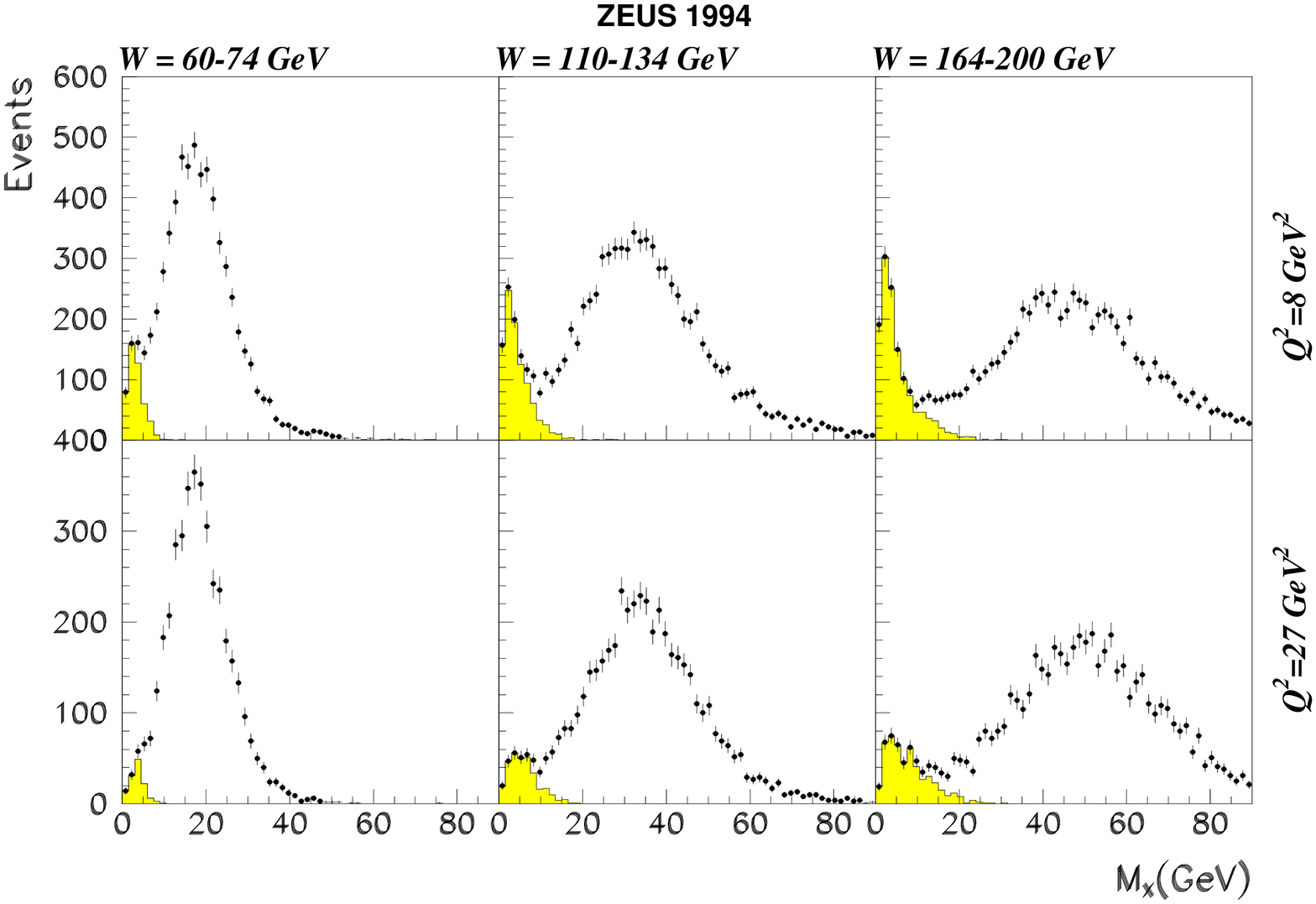,width=8.8cm,height=6.8cm}
\epsfig{file=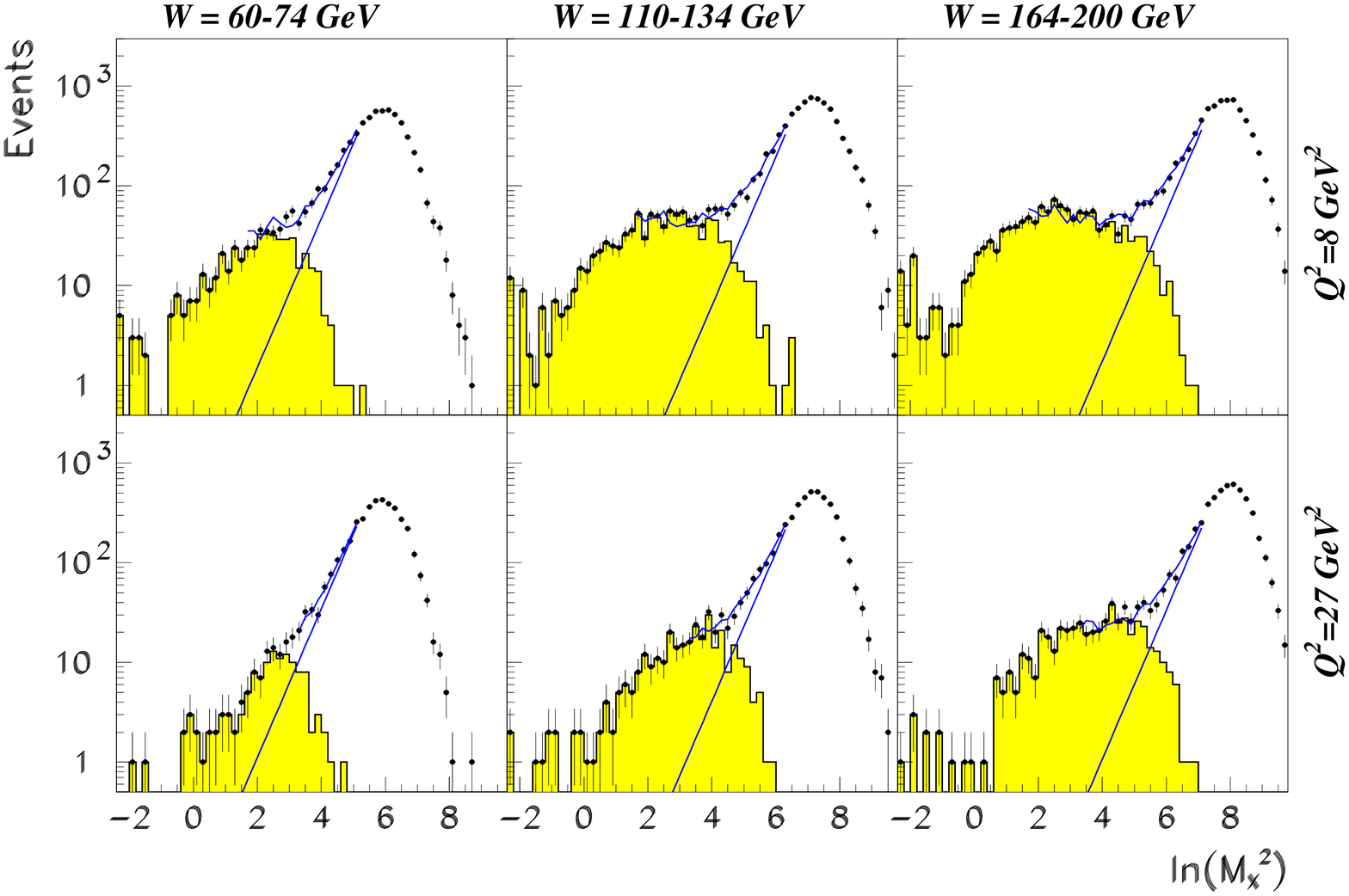,width=8.8cm,height=6.8cm}
\end{center}
\vspace{0.cm}
\caption{Distribution of $M_X$ (left) and $\ln M_X^2$ (right) for the ZEUS
1994 data.
The shaded histograms show the distribution of events with $\eta_{max} < 1.5$.
The straight lines in the right-hand side figures give the non-diffractive 
contributions as obtained from an exponential fit to the data.}
\label{fig:logmx}
\end{figure}

The ZEUS experiment \cite{ZEUS_mx,ZEUS_94,Kowalski_talk} has also exploited 
the fact that diffractive interactions are characterised by a non-exponentially 
suppressed rapidity gap.
% \cite{Bj_exp_suppressed}.
For fixed values of the hadronic centre of mass energy \W, this corresponds 
to a non-exponentially suppressed contribution to the \mx\ mass distribution 
at amall $M_X$, as reconstructed in the central detector.\footnote
{The pseudorapidity gap $\Delta \eta$ is related to $\ln M_X$ by the 
relation $\Delta \eta \approx \ \ln \ (W^2 / M_X^2)$.}
The diffractive contribution is observed in Fig. \ref{fig:logmx} for small 
values of \mx, whereas for large \mx\ the distribution can be parameterised 
as a single exponential (the falling flange at largest \mx\ values is due to 
detector acceptance).

%=======================\label{fig:logmx} =================
%====================\end{fig:logmx} ==================

\begin{figure}[tbp]
\vspace{0.cm}
\begin{center}
\epsfig{file=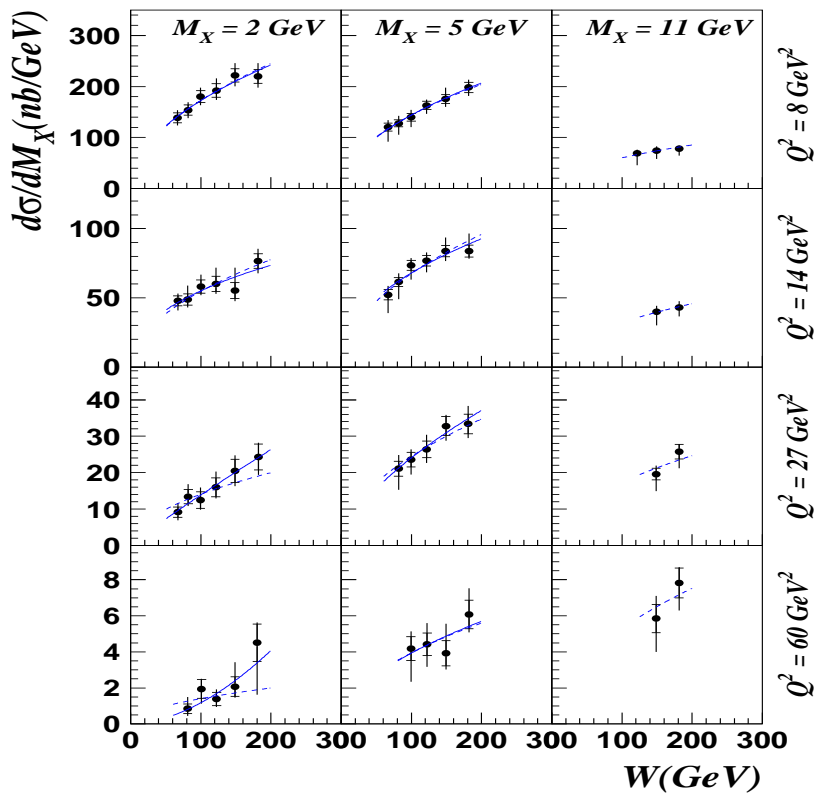,width=12cm,height=12cm}
\end{center}
\vspace{0.cm}
\caption{ZEUS measurement of the differential cross section 
${\rm d} \sigma^{diff}_{\gamma^*p \rightarrow XY} / {\rm d} M_X$
with $M_Y < 5.5$ \gev\ as a function of $W$ for several \mx\ and \qsq\ values.
The solide curves show the result of fitting the diffractive cross section
for each ($W$, \qsq) bin separately as a power of $W$; the dashed curves
show the result of the fit when the power of $W$ is assumed to be the same for
all bins.}
\label{fig:dsigmadmx}
\end{figure}

%The differential cross section 
%${\rm d} \sigma^{diff}_{\gamma^*p \rightarrow XN} / {\rm d} M_X$
%is presented in Fig. \ref{fig:dsigmadmx} as a function of $W$ for several
%\mx\ and \qsq\ values.
The diffractive cross section is thus extracted for several \qsq\    
and \W\ intervals (see Fig. \ref{fig:dsigmadmx}), after subtraction of the 
exponentially falling non-diffractive background.
The ZEUS detector acceptance implies that events are selected with $M_Y < 5.5$
\gev.
For the selected (\mx, $W$, \qsq ) bins, \xpom\ is effectively 
kept below 0.01, leading to a negligible contribution of reggeon exchange.

%===================\label{fig:dsigmadmx}===============
%====================\end{fig:ZEUS_f2d_94} ===============

%====================================================
%====================================================
%\subsection {LPS Measurement; the {\boldmath $t$} Slope} \label{sect:LPS}
\subsection {LPS Measurement; the t Slope} \label{sect:LPS}

Finally, the ZEUS collaboration has also measured the 
%\fdfourf \ 
diffractive structure function, using their leading proton spectrometer (LPS)
\cite{ZEUS_LPS}.

The use of proton spectrometers, which detect protons with energy 
close to the beam energy, provides a clean measurement of the 
diffractive cross section since the scattered proton is unambiguously 
tagged, avoiding contaminations due to proton dissociation events and to 
particle density fluctuations in the central detector.
Another advantage of the LPS is that the proton momentum measurement leads
to a direct determination of \xpom\ (see eq. \ref{eq:xpom_def}).
However, the drawbacks of the present ZEUS and H1 proton spectrometers are 
the limited statistics (the acceptance is $5 - 7 \%$) and the limited 
range in \ttt \ ($0.1 \lsim |t| \lsim 0.4$ \gevsq \ for $x_\pom < 0.03$), which 
implies that the cross section estimate requires an extrapolation in $t$ of 
the measurement.

\begin{figure}[tbp]
\vspace{0.cm}
\begin{center}
\epsfig{file=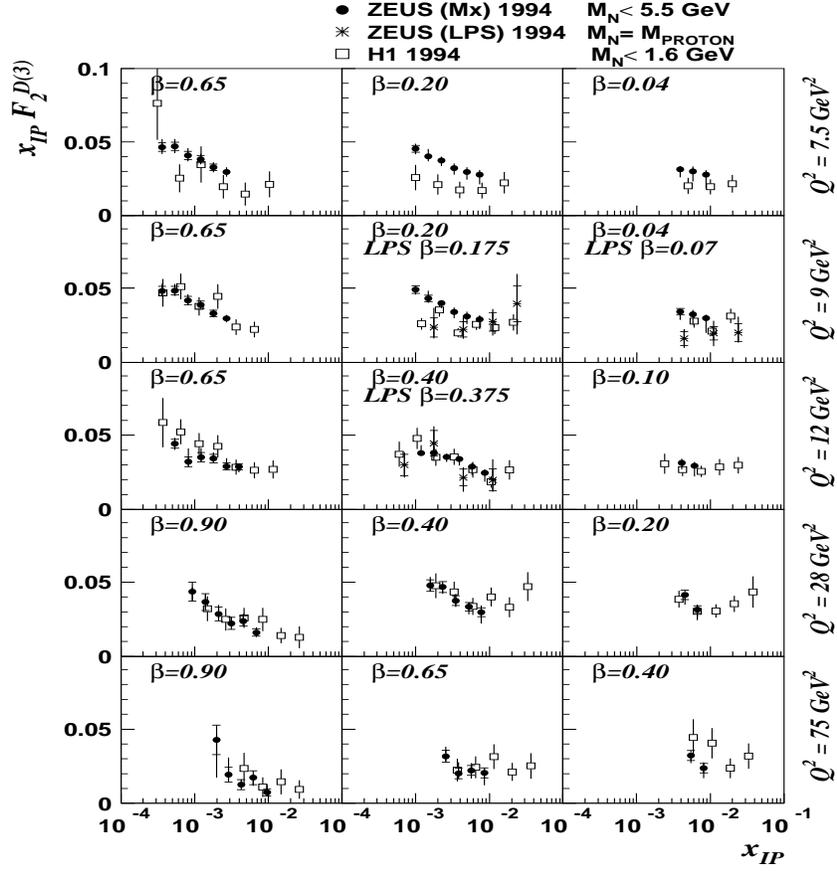,width=12cm,height=12cm}
\end{center}
\vspace{0.cm}
\caption{Measurement of $x_\pom\ \cdot $ \fdthreef\ as a function of 
$x_\pom$ for various \qsq\ and $\beta$ values with the 1994 data:
ZEUS LPS data (stars), ZEUS data with the $\ln M_X^2 $ method (solid points),
and a subsample of the H1 measurements (open squares).}
\label{fig:ZEUS_f2d_94}
\end{figure}

The diffractive structure function measured by ZEUS with the LPS is 
presented in Fig. \ref{fig:ZEUS_f2d_94}, together with the ZEUS $\ln M_X^2 $ 
and a subsample of the H1 measurements.
Given the large errors, the LPS measurement is in agreement with the other
results.

%===================\label{fig:ZEUS_f2d_94}===============
%====================\end{fig:ZEUS_f2d_94} ===============

Proton spectrometers have also the unique ability of providing a measurement of 
the $t$ distribution for inclusive events.
The ZEUS LPS has been used to measure the $t$ distribution both in photo- 
and electroproduction \cite{ZEUS_LPS,ZEUS_LPS_photoprod_t,ZEUS_LPS_t}.
For $3 < Q^2 < 150$ \gevsq , $x_\pom < 0.03$, $60 < W < 270$ \gev , 
$M_X > 2$ \gev\ and $0.073 < |t| < 0.40$ \gevsq, the $t$ distribution has 
been measured with the ZEUS LPS to be exponential, with a slope 
%$0.015 < \beta < 0.5$ and $x_\pom < 0.03$, the measured $t$ slope is 
% (see Fig. \ref{fig:tslope}):
%
\begin{equation}
%b = 7.2 \pm 1.1 {\rm \ (stat.) \ } ^{+0.7}_{-0.1} {\rm \ (syst.) } 
b = 7.1 \pm 1.0 {\rm \ (stat.)} \pm 1.2 {\rm \ (syst.) } 
                                                  {\rm \ GeV^{-2}}.
                                               \label{eq:tslope}
\end{equation}

%=======================\label{fig:tslope} =================
\begin{figure}[tbp]
\vspace{-1.cm}
\begin{center}
\epsfig{file=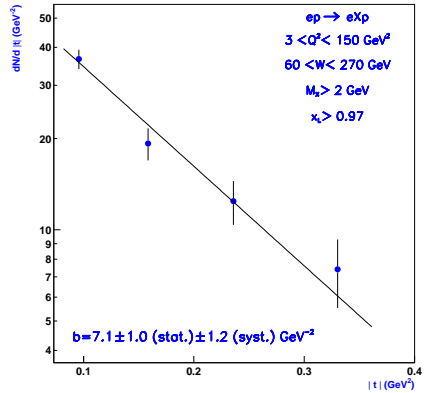,width=6cm,height=6cm}
\end{center}
\vspace{0.cm}
\caption{Differential $t$ distribution of diffractive interactions with
$3 < Q^2 < 150$ \gevsq , $x_\pom < 0.03$, $60 < W < 270$ \gev\ and 
$M_X > 2$, as measured by the ZEUS 
collaboration using the Leading Proton Spectrometer.}
\label{fig:tslope}
\end{figure}
%=======================\end{fig:tslope} ==================

No \qsq\ dependence of the slope is observed in the electroproduction data.
A photoproduction measurement gives a very similar value 
\cite{ZEUS_LPS_photoprod_t}.
This has to be contrasted with the strong dependence with \qsq\ of the
$\rho$ meson production slope.

%====================================================
%====================================================
\subsection {``Soft'' and ``Hard'' Inclusive Diffraction} 
                                                      \label{sect:soft_hard}

Although ``soft'' diffraction may be only an ``effective'' concept, 
%\cite{soft_hard}, 
it will be used here as referring to a mild energy 
dependence in the parameterisation of eq. \ref {eq:en_dep}, typical of 
hadron$-$hadron scattering at high energy (cf. eq. \ref{eq:traj_pom}), 
in contrast with 
the stronger energy dependence characterised by a pomeron intercept of the 
order of $1.2 - 1.3$, as observed in inclusive DIS \cite{H1_f2,ZEUS_f2} 
and in diffractive \jpsi\ production \cite{jpsi,jpsi2}.

In photoproduction, measurements of the inclusive diffractive cross section 
have been performed by the H1 \cite{h1_photoprod} 
and ZEUS \cite{zeus_photoprod} experiments.
In both cases, the energy dependence was found to be consistent with ``soft'' 
diffraction.
A triple-Regge analysis was performed of the $W$ and \mx\ dependences of the
H1 data, complemented with lower energy data.
While this analysis indicates the presence of a sizeable non-diffractive 
contribution, the fitted pomeron intercept is 
\begin{equation}
   \alpha_\pom(0) = 1.07 \pm 0.05.
   				 \label{eq:alphapom_photoprod_H1}
\end{equation}				
The description of the ZEUS data in the range $8 <M_X < 24$ \gev\ with a purely
triple-pomeron diagram gives for the pomeron intercept
\begin{equation}
   \alpha_\pom(0) = 1.12 \pm 0.09;
   				 \label{eq:alphapom_photoprod_ZEUS}
\end{equation}				
a substantial non-diffractive contribution is found to be necessary at low
mass.

\begin{figure}[tbp]
\vspace{0.cm}
\begin{center}
\epsfig{file=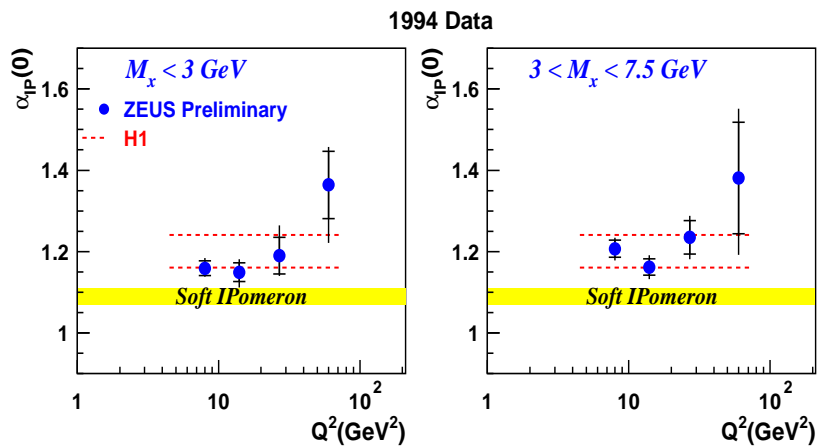,width=12cm,height=6cm}
\end{center}
\vspace{0.cm}
\caption{Measurement of $\alpha_\pom(0)$ for H1 (the measurement limits are 
indicated as the dotted lines) and for ZEUS (dots).}
%\cite{??}.}
\label{fig:alphapom}
\end{figure}

In contrast, in DIS diffractive scattering, the pomeron intercept 
$\alpha_{\pom}(0)$ is inconsistent with a ``soft'' value (see Fig. 
\ref{fig:alphapom}). 
The H1 measurement is
\begin{equation}
   \alpha_\pom(0) = 1.20 \pm 0.02 {\rm \ (stat.)} \pm 0.01 {\rm \ (syst.)}
                        \pm 0.03 {\rm \ (model)},
						\label{eq:alphapom_DIS_H1}
\end{equation}				
with no significant $\beta$  or \qsq \ dependence over the range 
$0.4 < Q^2 < 75$ \gevsq\ \cite{H1_f2d_94,H1_f2d_SV}.
The ZEUS measurements are similar, also with no significant \qsq\ dependence 
over the measured range
\cite{ZEUS_94,Kowalski_talk} (see Fig. \ref{fig:alphapom}).
Both measurements lie significantly above the ``soft'' pomeron values,
although they seem to be below the ``hard'' values measured in inclusive DIS.
This suggests that inclusive diffraction may be putting the stage
for interplay between soft and hard processes.

%=======================\label{fig:alphapom} =================
%=======================\end{fig:alphapom} =================

%However, they are still below the ``hard'' value of about 1.3.

%Similarly, the slope of the $t$ distribution in diffractive DIS at HERA 
%(cf. section \ref{sect:LPS}) is ``harder'' than that measured in a typical 
%soft process: $b \simeq\ 10-11 {\rm \ GeV^{-2}}$ for $\rho$ photoproduction 
%at small $|t|$ \cite{h1_zeus_rho_photoprod}, but it is higher than in a
%typical hard process: 
%$b \simeq 4-5 {\rm \ GeV^{-2}}$ for \jpsi\ photoproduction \cite{jpsi}.
%The slope measured for inclusive diffraction is thus similar to the slope 
%$b \simeq 6-8 {\rm \ GeV^{-2}}$ measured  for $\rho$ electroproduction 
%in the transition domain with \qsq\ $ \simeq 10$ \gevsq.
%
%These results thus consistently suggest that inclusive diffraction is 
%subjected to an interplay between soft and hard processes.

%====================================================
%====================================================
\section {Partonic Structure of Diffraction}
      							\label{sect:parton}
							
Given the fundamental aspect of diffraction in hadron$-$hadron interactions,
its understanding in terms of partons is a major challenge for QCD.
This understanding is helped by the use of two complementary pictures: photon 
hadronic fluctuations and pomeron structure function.
The parton densities in the pomeron extracted from the study of scaling
violations can then be extended, under the assumption of factorisation,
to the study of inclusive final states and semi-inclusive processes.

%====================================================
%====================================================
\subsection {Two Complementary Pictures}
      							\label{sect:two_pict}

In a traditional approach of diffraction, LRG events are attributed to 
the exchange in the $t$-channel of a colourless object, the pomeron, which 
is sensitive to hard interactions, is flavour blind and carries the quantum 
numbers $J^{PC}I^G = 0^{++}0^+$ of the vacuum. 
In this context, diffractive interactions can be conveniently visualised in 
two different Lorentz frames.

\begin{figure}[tbp]
\vspace{-1.cm}
\begin{center}
\epsfig{file=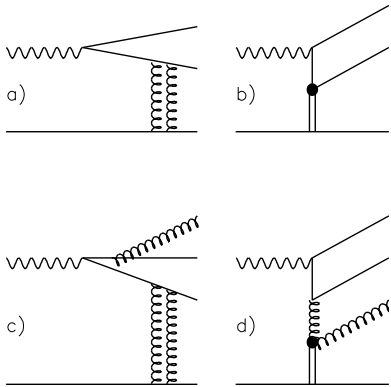,width=7cm,height=7cm}
\end{center}
\vspace{-1.cm}
\caption{Photon-proton diffractive interaction: a,c) visualised in a frame 
moving fast with respect to the photon: the photon fluctuates in a 
$q \bar{q}$ (a) or a $q \bar{q} g$ Fock state (c), wich subsequently 
diffractively scatters on the proton (here, the pomeron is modelised as a
two-gluon system, and only one of the relevant diagrams is shown); 
b,d) visualised in the Breit frame, the pomeron being parameterised as a 
$q \bar{q}$ system, with direct photon$-$quark coupling (b), or as a $gg$
system, with photon-gluon fusion (d).}
\label{fig:topo}
\end{figure}

In the {\it proton rest frame} (see Fig. \ref{fig:topo}a,c), or any fast 
moving frame with respect to the photon, relativistic time dilatation implies 
that photon fluctuations into hadronic systems 
($q \bar{q}$, $q \bar{q} g$, etc.), taking place at a long distance from 
the proton, appear 
as ``frozen'' during the (much shorter) interaction time with the proton.
In this approach, the diffractive cross section is thus calculated as the 
convolution of three factors, corresponding respectively to the (long-lived) 
photon hadronic structure, the (short time) diffractive interaction between 
the photon hadronic components and the proton, and the (long-time) 
hadronisation and final state parton recombination:
\begin{equation}
\sigma_D (\gamma + p \rightarrow X + Y)
 = \sum_{q \bar{q},q \bar{q} g, ...}
       \int {\rm d}^2 b_T \
       \Psi (\gamma \rightarrow  q \bar{q},...) \cdot
       \sigma (q \bar{q},... + p \rightarrow q \bar{q},... + p) \cdot
       \Psi (q \bar{q},... \rightarrow hadrons).
							\label{eq:convol}
\end{equation}
These calculations are usually performed in the impact parameter space ($b_T$), 
and only the first Fock states ($q \bar{q}$, $q \bar{q} g$) are
considered.
%\cite{Fock_states}.
The pomeron is parameterised as a two gluon sytem or a Lipatov ladder, which
meets the requirements of colour neutrality and flavour blindness.

The {\it pomeron structure function} approach is developed in the pomeron 
(or proton) infinite momentum frame, and views the pomeron 
as a colour singlet partonic object emitted from the proton.
In this approach, if the concept of flux factorisation {\it \`{a} la}
Ingelman-Schlein \cite{Ingelman-Schlein} holds, the 
virtual photon probes the pomeron structure similarly to the probing of the 
proton structure by the photon in ``normal'' DIS.\footnote
{Although factorisation is not expected to hold when transporting parton
distributions extracted in DIS diffraction to the case of hadron$-$hadron 
interactions or of interactions with a resolved photon \cite{NNN92,non_fact}, 
it was shown to hold for diffractive DIS and direct photoproduction 
\cite{fact}.}
The DGLAP equations may then describe the QCD evolution of the pomeron 
structure function (see however below, section \ref{sect:pdf}).

%=======================\label{fig:topo} ==================
%=======================\end{fig:topo}====================

These two complementary pictures of diffraction have of course to be consistent 
with each other. 
In a semiclassical approach, they have been shown to be equivalent in the
case of soft gluon emission off the photon 
(a $q \bar{q} + $ soft $g$ Fock state) \cite{Hebecker_link}.

%======================================================
%======================================================
\subsection{Alternative Approaches}                        \label{sect:SCI}

In recent years, models have been proposed to explain the production of LRG
events in excess over expectations for particle fluctuation, without 
reference to the specific concepts of diffraction or pomeron exchange.
Here, the formation of LRG events is seen as a two-step process. 
The first step is similar to ``normal'' DIS events, a quark being
ejectecd off the proton by the photon.
In a second stage, soft colour interactions (SCI) modify 
the colour properties of the outgoing system, leading to the formation of 
two colour-neutral systems separated by a gap in rapidity.

In a semi-classical approach \cite{Buch}, the propagation
of the struck parton through the proton colour field is accompanied by soft 
colour rotation which leads, for a fraction of the events, to colour 
neutralisation.

The concept of SCI has also been implemented into the LEPTO Monte Carlo 
calculation \cite{LEPTO5.1} where, before hadronisation, parton 
reconnection takes place through the exchange of soft gluons, leading to 
the formation of colour neutral systems.

%======================================================
%======================================================
\subsection {Parton Distributions in the Pomeron}
							\label{sect:pdf}

\begin{figure}[tbp]
\vspace{0.cm}
\begin{center}
\epsfig{file=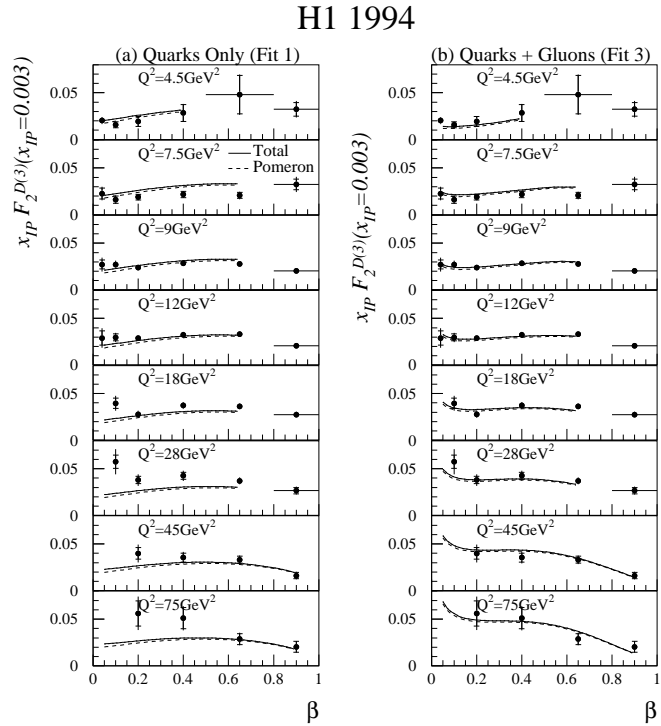,width=10cm,height=10cm}
\end{center}
\vspace{0.cm}
\caption{H1 measurement (1994 data) of $x_\pom\ \cdot $ \fdthreef \  
interpolated to $x_\pom = 0.003$, as a function of $\beta$ for several \qsq\ 
values.
The superimposed curves represent a) a DGLAP fit with quarks only at the 
starting scale $Q_0^2 = 3 $ \gevsq ; b) the preferred QCD fit with quarks and 
gluons at the starting scale.} 
\label{fig:scal_viol_beta}
\end{figure}

\begin{figure}[tbp]
\vspace{0.cm}
\begin{center}
\epsfig{file=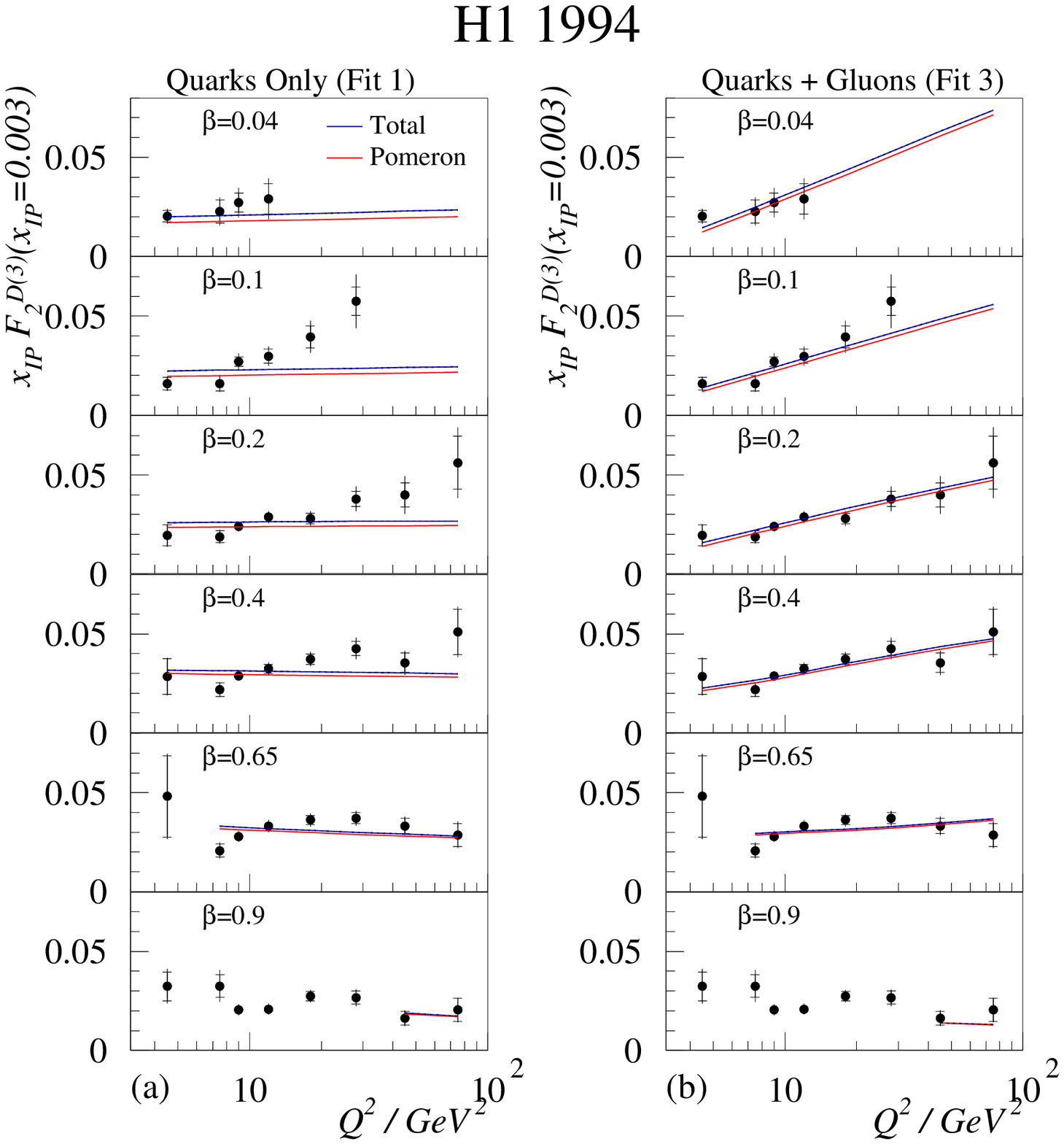,width=10cm,height=10cm}
\end{center}
\vspace{0.cm}
\caption{H1 measurement (1994 data) of $x_\pom\ \cdot $ \fdthreef \ 
interpolated to $x_\pom = 0.003$, as a function of \qsq\ for several 
$\beta$ values. 
The superimposed curves represent a) a DGLAP fit with quarks only at the 
starting scale $Q_0^2 = 3 $ \gevsq ; b) the preferred QCD fit with quarks and 
gluons at the starting scale.}
\label{fig:scal_viol_qsq}
\end{figure}

Fig. \ref{fig:scal_viol_beta} presents the 1994 H1 measurement of 
$x_\pom\ \cdot$ \fdthreef\ interpolated to $x_\pom = 0.003$,\footnote
{At this low \xpom\ value, the reggeon contribution to \fdthree\ is negligible.}
as a function of $\beta$ for several \qsq\ values.
In strong contrast to the hadron structure function at high $x$, the pomeron
structure function is large for high $\beta$ values, even for relatively large
\qsq.
In Fig. \ref{fig:scal_viol_qsq}, the H1 measurement is shown as a function 
of \qsq\ for several $\beta$ values.
Scaling violations with a positive slope are observed up to large $\beta$ 
values, which suggests that the pomeron is dominated by hard gluons.

%=======================\label{fig:scal_viol_beta} ================
%=======================\end{fig:scal_viol_beta} ==================

%=======================\label{fig:scal_viol_qsq} ================
%=======================\end{fig:scal_viol} ==================

To analyse quantitatively the pomeron structure, DGLAP fits to the 1994
diffractive structure function presented in Fig. \ref{fig:H1_f2d_94} 
have been performed by H1 for the two contributions of eq. \ref{eq:factoris2},
thus assuming factorisation \cite{H1_f2d_94}.\footnote
{Including the 1995 shifted vertex low \qsq\ data does not change the results of
the analysis \cite{H1_f2d_SV}.} 
The reggeon contribution is parameterised using the pion 
structure function \cite{Owens_pi}, and two cases are considered for the
pomeron structure function.
In the first case, only quarks are allowed to contribute at the starting scale 
$Q_0^2 = 3 $ \gevsq\ (``fit 1'' in \cite{H1_f2d_94}); 
this gives a poor $\chi^2$: 314 for 159 d.o.f. $-$ see the 
curves superimposed on Figs. \ref{fig:scal_viol_beta}a, \ref{fig:scal_viol_qsq}a.
In the second case, gluons are also allowed to contribute at the starting 
scale, and a good description of the data is obtained: 
$\chi^2 / {\rm d.o.f.} = 176/154$ $-$ see the curves on Figs. 
\ref{fig:scal_viol_beta}b, \ref{fig:scal_viol_qsq}b.

\begin{figure}[tbp]
\vspace{-0.5cm}
\begin{center}
\epsfig{file=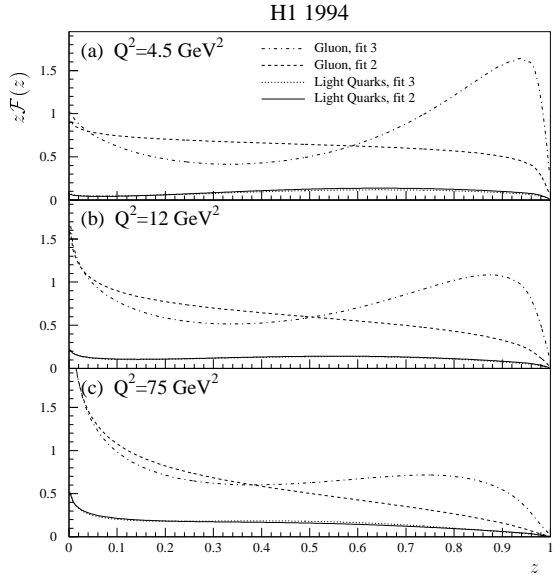,width=8cm,height=8cm}
\end{center}
\vspace{0.cm}
\caption{Parton distributions for three \qsq\ values, as obtained from 
DGLAP fits to the 1994 H1 measurement of $x_\pom\ \cdot$ \fdthreef,
as a function of $z$, the pomeron momentum fraction carried by the parton
entering the hard interaction. 
The curves represent the gluon contributions for the best fit (``fit 3'', 
dashed-dotted curves) and for the case of a flat gluon contribution at
the starting scale (``fit 2'', dashed curves); the lower curves correspond
to the light quark contributions.}
\label{fig:parton_distr}
\end{figure}

From this DGLAP analysis, parton distributions are extracted.
The distributions corresponding to the best fit to the H1 data 
(``fit 3'' in \cite{H1_f2d_94}) are presented in 
Fig. \ref{fig:parton_distr} as a function of $z$, the pomeron 
momentum fraction carried by the parton entering the hard interaction 
(note that $\beta= z$ for quarks but $\beta \leq z$ for gluons). 
This distribution shows an enhancement of the gluon contribution at high $z$
for small \qsq\ values (the turn-over at high $z$ was forced into the fit
in order to avoid a singular behaviour of the gluon distribution).
A nearly equally good fit (``fit 2'' in \cite{H1_f2d_94}), also presented
in Fig. \ref{fig:parton_distr}, is obtained with 
a gluon distribution flat in $z$ at the starting scale.
In both cases, the gluons amount to $\geq 80 \%$ of the pomeron partonic
content in the present \qsq\ range, and the parton distribution functions are
dominated by hard gluons.

Similar studies were performed by the ZEUS collaboration \cite{ZEUS_fit}
(see also \cite{jim_fit}).
In this case, simultaneous fits were performed to the \fdthreef\ 
distributions in DIS and to the jet diffractive photoprodution cross section.
These analyses also lead to the conclusion that gluons dominate the pomeron 
structure: 
at a scale of 4 \gevsq, the ZEUS analysis indicates that the fraction of the
pomeron momentum carried by gluons lies between 0.64 and 0.94. 
Note however that in such a simultaneous fit,
%the assumption is made that 
%the same ``pomeron'' is at work for inclusive diffractive interactions and 
%for hard jet production: 
possible factorisation breaking effects (e.g. related to secondary 
interactions of photon remnants with the proton in the resolved regime of
photoproduction) are not taken into account, nor a possibly different 
interplay between ``hard'' and ``soft'' diffraction for the two processes. 

From an empirical point of view, a major merit of these DGLAP analyses is that 
they provide good fits to the data, based on simple assumptions, and that
the results can easily be implemented in Monte Carlo simulations in
order to test in various semi-inclusive analyses, the universality of the 
extracted parton distributions.

%=======================\label{fig:parton_distr}=================
%=======================\end{fig:parton_distr} ================

\begin{figure}[tbp]
\vspace{0.cm}
\begin{center}
\epsfig{file=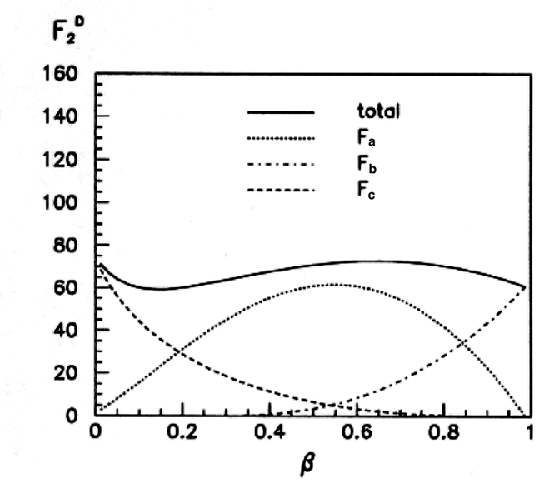,width=6cm,height=6cm}
\end{center}
\vspace{0.cm}
\caption{Shape of the three contributions expected to contribute to 
inclusive DIS diffraction: $q \bar{q} g$ Fock states at small $\beta$ values
(dashed line, ${\rm F_c}$); transversely polarised $q \bar{q}$ states at 
medium $\beta$ values (dotted line, ${\rm F_a}$); longitudinally polarised 
$q \bar{q} g$ states at large $\beta$, with a higher twist behaviour 
(dash-dotted line, ${\rm F_b}$).}
\label{fig:evol}
\end{figure}

It is important to note however that the use of the DGLAP evolution equation 
may not be valid over the whole $\beta$ range.
It has been stressed (see e.g. \cite{NNN92,evol}) that three different 
contributions, with
different \qsq\ evolutions, may dominate different $\beta$ regions: 
$q \bar{q} g$ at small $\beta$ (large $M_X$ masses);
% and to obey a DGLAP evolution (ref. NZ ??); 
transversely polarised $q \bar{q}$ systems in the central $\beta$ region 
($0.2 \lsim \beta \lsim 0.8)$;
%(DGLAP-type??) leading twist contribution;
longitudinally polarised $q \bar{q}$ systems for large $\beta$ values, giving 
a higher twist contribution.
The importance of the longitudinal contribution at high $\beta$ is supported by 
the observation of a dominant longitudinal cross section for vector meson 
production (following eq. \ref{eq:beta_def}, low mass vector mesons are
generally produced with large $\beta$ values). 
A simple DGLAP analysis through the whole kinematic domain can thus be 
dubious,
especially since the superposition of the three contributions could mimic the
distributions in Fig. \ref{fig:scal_viol_beta} 
(see Fig. \ref{fig:evol}, taken from \cite{evol}).
However, it should be noted that the H1 conclusions are basically
unaffected if the fit domain is restricted to $\beta < 0.65$.\footnote
{Since the Rio Workshop, a model -- of which the general lines are based on
an analysis of jet diffractive production \cite{Bartels_rio} -- has been 
proposed to describe inclusive
diffraction \cite{BEKW}.
This model provides a parameterisation of the three major 
contributions (leading twist longitudinal and transverse cross sections, and 
higher twist longitudinal contribution). 
This model has been compared to the ZEUS \cite{ZEUS_94} and H1 \cite{H1_f2d_SV}
measurements.  
The two experiments give different numerical results for the fitted
parameters, which can be traced to 
the differences in the measurements visible in Fig. \ref{fig:ZEUS_f2d_94}. 
More precise measurements and a better understanding of systematic 
uncertainties are thus necessary before a detailed interpretation of the fit 
results is possible. 
However, a common feature for both experiments is the need for a significant 
contribution of the higher twist longitudinal cross section at high $\beta$.
Two models based on BFKL dynamics \cite{dipole,NNN} were also compared to the 
H1 measurement \cite{H1_f2d_SV}.}

%====================\label{fig:evol}==================
%====================\end{fig:evol} ==================

%=======================================================
%=======================================================
\subsection {Monte-Carlo Programs: RAPGAP and LEPTO}
							\label{sect:MC}

Monte-Carlo programs provide basic tools for experimental studies: they are
used to correct the raw data for acceptance, efficiency and smearing effects, 
and modelise theoretical predictions in a form directly comparable to 
the measurements.

In the field of diffraction, two main Monte-Carlo simulations are 
available for DIS interactions: the RAPGAP and LEPTO programs.

The RAPGAP program \cite{rapgap}, based on a factorisable pomeron flux, uses parton 
distribution functions in the pomeron evolved following the DGLAP equations.
At the starting scale, the distributions can be modelised as a $q \bar{q}$ 
system, following the results of the fits to the \fdthree\ scaling violations,
or any other form.
The photon interaction thus takes place directly on a quark in the pomeron 
(Fig. \ref{fig:topo}b), or via boson-gluon fusion (Fig. \ref{fig:topo}d).
In both cases, a ``pomeron remnant'' is present (a quark in the first case,
a gluon in the second case).
Higher order contributions (e.g. the QCD-Compton diagram), hadronisation 
processes and QED radiation are incorporated.

The LEPTO 5.1 Monte Carlo program \cite{LEPTO5.1} is the implementation of the 
concept of SCI in the framework of the Lund model. 
LRG event are produced as for ``normal'' deep-inelastic scattering, using 
standard parameterisations of the parton densities in the proton, 
but parton reconnection by soft gluons, with pure colour exchange and
no kinematics modification, generates neutral partonic systems. 
Compared to ``normal'' DIS interactions, the only adjustable parameter in LEPTO 
5.1 is the amount of soft colour interactions, and the program is thus highly 
constrained.

For photoproduction, the POMPYT program \cite{pompyt} is a diffractive-specific 
extension to PYTHIA, containing both direct and resolved photon interactions.

%=======================\label{fig:rapgap}==================
%\begin{figure}[tbp]
%\vspace{1.cm}
%\begin{center}
%\epsfig{file=fig_diaga.eps,width=6cm,height=6cm}
%\epsfig{file=fig_diagb.eps,width=6cm,height=6cm}
%\end{center}
%\vspace{1.cm}
%\caption{Photon-pomeron interaction in the RAPGAP Monte-Carlo model: 
%a) $q \bar{q}$ modelisation of the pomeron; b) $g g$ modelisation.	}
%\label{fig:rapgap}
%\end{figure}
%=======================\end{fig:rapgap} ==================

%=========================================================
%=========================================================
\section{Inclusive and Semi-inclusive Final State Studies}
							\label{sect:final_state}

The H1 and ZEUS collaborations have performed numerous studies of the features 
of inclusive and semi-inclusive diffractive final states (the photon dissociation 
system $X$).
The most inclusive studies concern the event shape (thrust
% \cite{H1_thrust} 
and sphericity),
% \cite{ZEUS_thrust}), 
charged particle multiplicities (total multiplicity, 
rapidity distributions, forward-backward correlations),
% \cite{H1_mult}), 
energy flow and single particle properties ($x_F$ distributions,
transverse momentum spectra, the ``sea-gull'' plot).
%\cite{H1_final_state,ZEUS_final_state}).
Semi-inclusive studies have been performed of jet 
%\cite{H1_jets,ZEUS_jets,ZEUS_jet_gap,H1_jet_gap,H1_high_t} 
and of charm production. 
%\cite{H1_charm,ZEUS_charm}.
These measurements are reviewed in detail in separate contributions to this 
Conference \cite {Valkarova,Zsembery,Cox}, and only motivations and 
general features of the analyses will be presented in this introduction.

Although definite theoretical predictions are presently lacking, especially 
for the overall features of diffractive final states, an experimental study 
of the data has provided useful informations on the structure of diffraction.
This information was gained both through model independent comparisons of the
characteristics of diffractive events with those of other processes, and 
through the confrontation to the data of Monte Carlo predictions.
In particular, RAPGAP calculations are used to test the sensitivity of the 
final state features to the input parton distributions, as obtained 
from DGLAP fits to the total inclusive diffractive cross section 
(see section \ref{sect:pdf}).
The pomeron modelisation as a $q \bar q$ system, although inconsistent with 
the fits to the structure function measurements, is often used for comparison.

%=========================================================
\subsection {Expected Qualitative Features of Diffractive Final States}
						\label{sect:qual_features}

It is useful to contrast, as a first order approach, the implications for the 
hadronic final state of two basic underlying parton topologies: two- and 
three-body final states, as illustrated in Fig. \ref{fig:topo}.

The left-hand side diagrams (a,c) of Fig. \ref{fig:topo} correspond to the 
photon fluctuation picture (the strong interaction between the proton and the 
photon hadronic fluctuations, modelised in the simplest case as two-gluon 
exchange, is illustrated by one of the relevant diagrams).
The right-hand side diagrams (b,d) illustrate the pomeron structure function 
approach.
 
The upper two figures correspond to two-body final states:
the $q \bar q$ Fock state of the photon (a), and the $q \bar q$ contribution
to the pomeron (b).

The lower two figures correspond to three-body final states.
The left-hand side diagram (c) corresponds to the $q \bar q g$ Fock state of 
the photon.
In the right-hand side picture (d), the pomeron is a two-gluon object, which 
interacts with the photon through boson-gluon fusion (BGF).

The characteristic feature of the two-body case is a jetty structure of 
the $X$ system, aligned with the photon-pomeron direction.
Photon fluctuations into $q \bar q$ pairs with large transverse momenta 
(Fig. \ref{fig:topo}a) have small interaction cross section with the proton
because this topology corresponds, through the uncertainty principle, to a small 
transverse distance between the quarks, which thus screen each other and 
form a nearly colour neutral system when seen from the proton.
This screening effect damps the large $p_T$ contributions, and favours 
an aligned jet topology.
Note that, in the structure function approach for a $q \bar q$ pomeron 
(Fig. \ref{fig:topo}b), QCD-Compton emission can induce an increase of the 
final state sphericity and the generation of large transverse momenta, 
but this process is suppressed by an additional power of $\alpha_s$.

Three-body final states are characterised by a dominant effective colour 
octet-octet interaction \cite{evol}.
In the $q \bar q g$ Fock state picture of Fig. \ref{fig:topo}c, both the fast 
$q \bar q$ system and the soft gluon are colour octets.
The topology is similar for Fig. \ref{fig:topo}d, with a forward going 
$q \bar q$ system and a pomeron remnant.
The octet-octet interaction between the forward and backward regions in Fig.
\ref{fig:topo}c,d leads to an increased activity (energy flow, particle 
multiplicity) in the central region, compared to the triplet-triplet case of 
the two body $q \bar q$ final states of Fig. \ref{fig:topo}a,b, 
expected to be close to the $e^+e^-$ case.

%=========================================================
%=========================================================
\subsection {Inclusive Final States}
							\label{sect:FS}

\begin{figure}[tbp]
\vspace{.cm}
\begin{center}
\epsfig{file=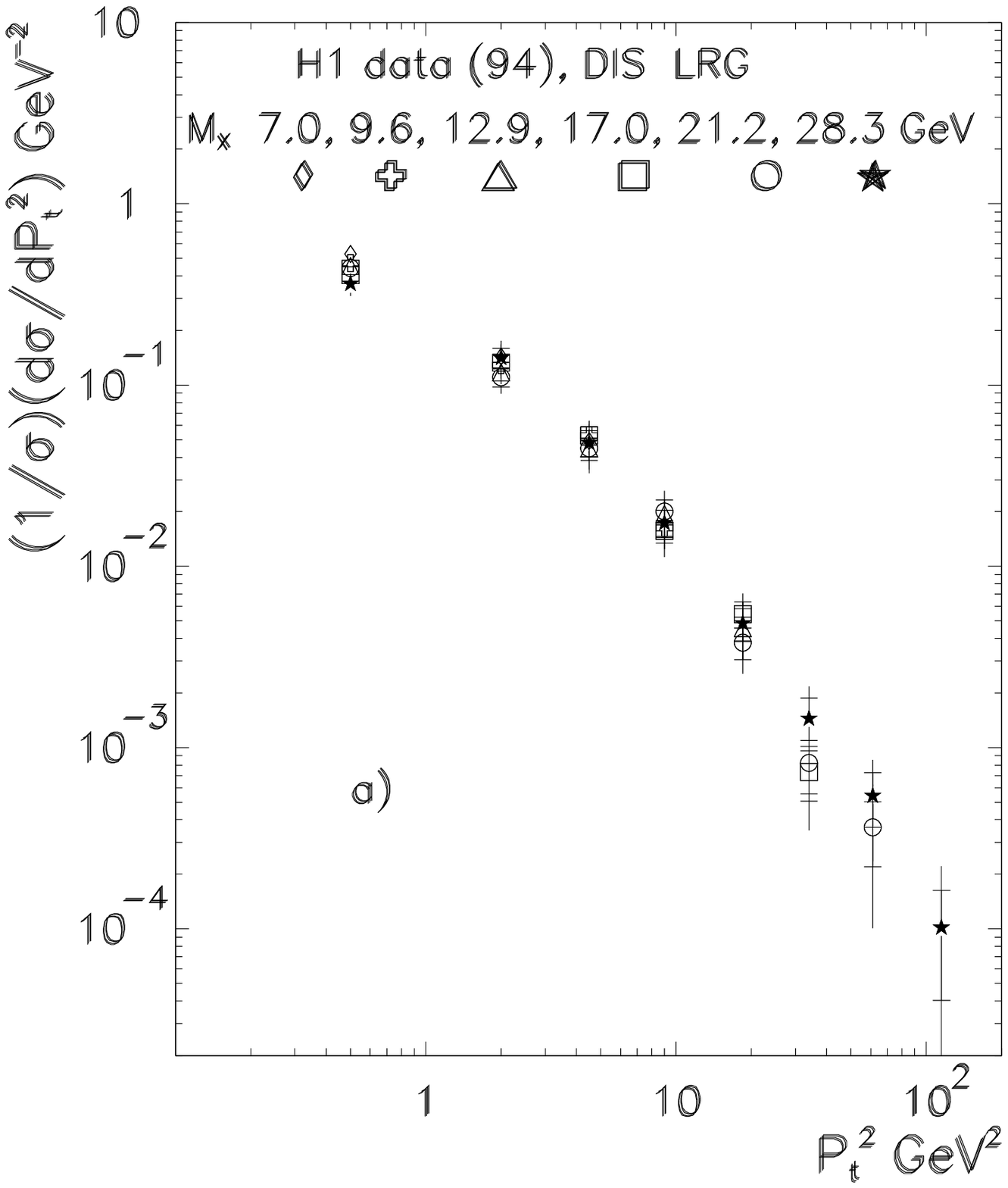,width=8cm,height=8cm}
\epsfig{file=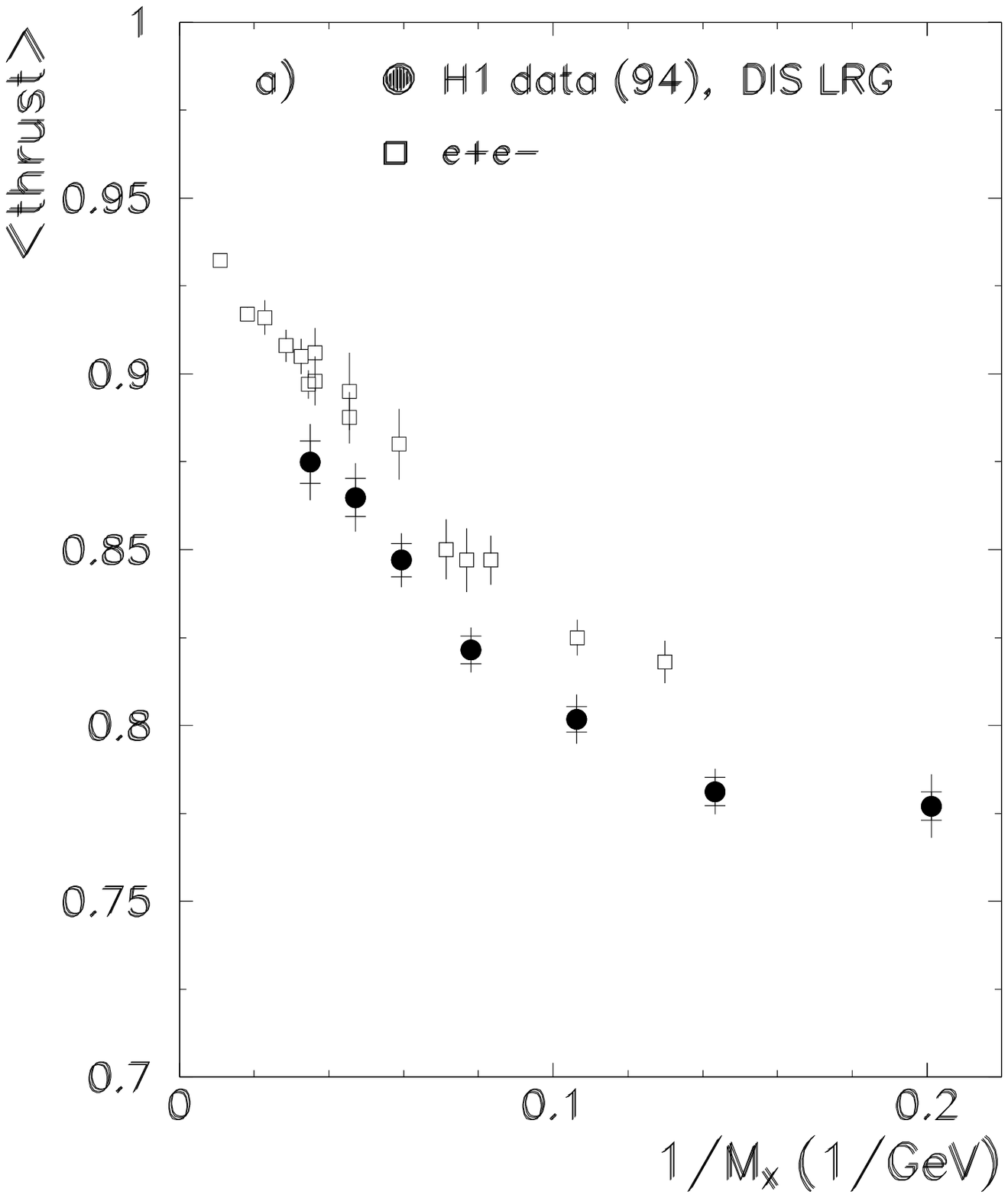,width=8cm,height=8cm}
\end{center}
\vspace{.cm}
\caption{H1 measurement with the 1994 data of 
a) the normalised thrust jet $P^2_t$ distributions for six $M_X$ intervals;
b) the average thrust, as a function of $1 / M_X$ (solid circles); the open
squares are for $e^+e^-$ data, with $s(e^+e^-) = M_X^2$.}
%, as a function of $1 / \sqrt{s_{e^+e^-}}$.}
\label{fig:H1_thrust}
\end{figure}

The event shape of diffractive events has been studied 
by H1 \cite{H1_thrust} and ZEUS \cite{ZEUS_thrust,ZEUS_final_state}.
The $X$ system is mostly aligned with the photon-pomeron direction.
However, a significant fraction of the events have a large $P_t$,
the component of the thrust jet momentum transverse to the photon direction 
in the $X$ system rest frame (see Fig. \ref{fig:H1_thrust}a).
In addition, for the same kinematic domain, the average thrust value is 
smaller than for $e^+ e^-$ interactions, and the sphericity is larger 
(see Fig. \ref{fig:H1_thrust}b).\footnote
{Although the LPS ZEUS data \cite{ZEUS_final_state} are compatible with the
$e^+ e^-$ results, they are affected by large errors, and are also compatible 
with the LRG results of H1 \cite{H1_thrust} and ZEUS \cite{ZEUS_thrust}.}
These features suggest that, compared to a basic aligned jet two-parton 
topology, which is expected to be close to the topology observed in
$e^+e^-$ interactions, higher parton multiplicities are also at work 
in diffraction.
This is confirmed by the comparison with RAPGAP predictions: thrust values 
for a purely $q \bar q$ pomeron are significantly higher than observed, 
whereas the gross features of the data are reasonably reproduced by a gluon 
dominated pomeron, following the parton distributions obtained from the 
inclusive cross section measurement.

%=======================\label{fig:H1_thrust} =================
%=======================\end{fig:H1_thrust} =================

\begin{figure}[tbp]
\begin{center}
\vspace{0.cm}
  \setlength{\unitlength}{1.0cm}
  \begin{picture}(16.0,12.0)
    \put(0.0,0.0){\epsfig{file=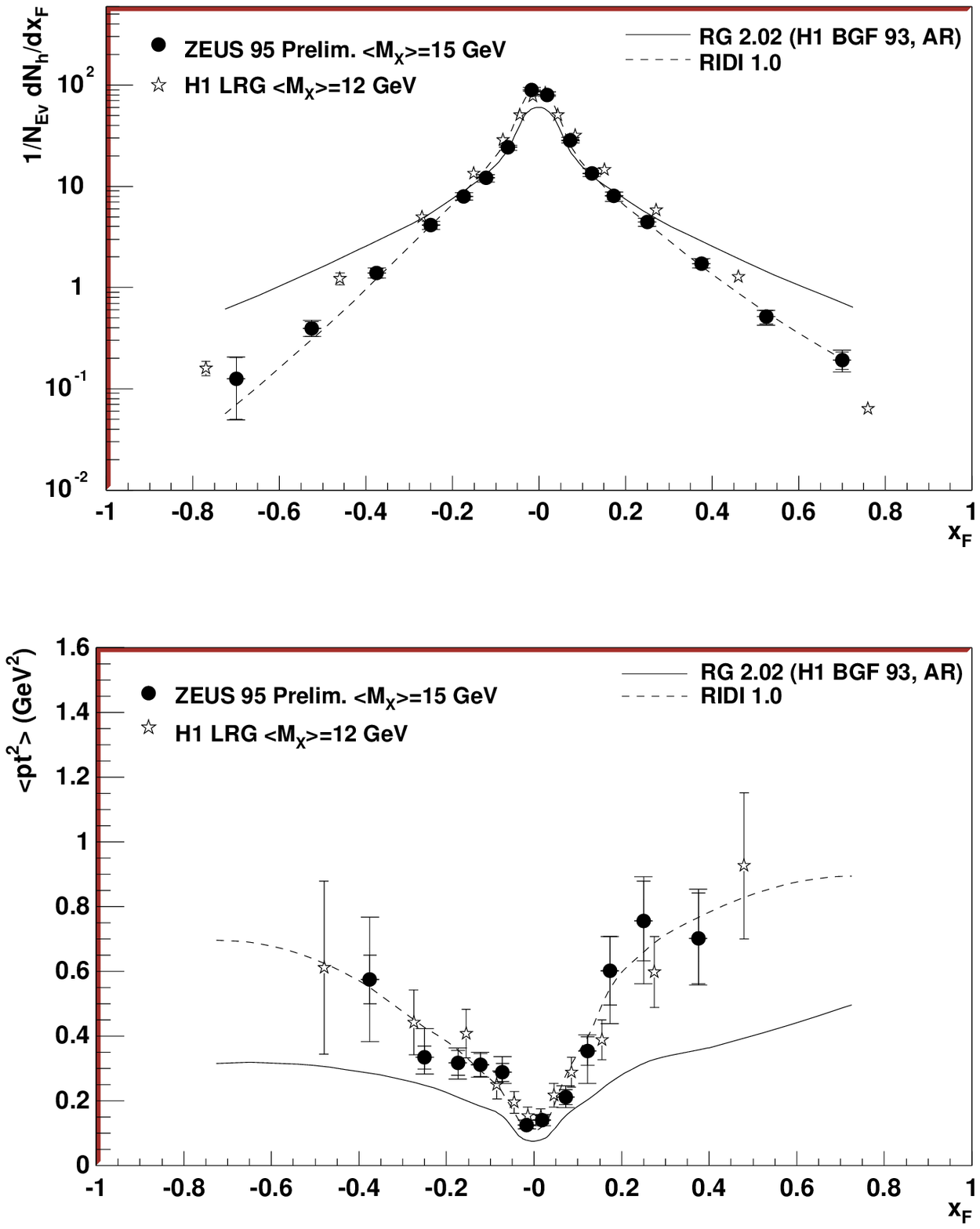,width=8cm,height=12cm}}
    \put(8.0,4.9){\epsfig{file=zeus.van.part_spectra.eps,width=8cm,height=12cm}}
    \put(0.0,0.0){\epsfig{file=whitebox.eps,width=8cm,height=6cm}}
    \put(8.0,10.5){\epsfig{file=whitebox.eps,width=8cm,height=6cm}}
  \end{picture}
\end{center}
\vspace{-6.0cm}
\caption{Measurements with the ZEUS LPS (closed circles) and by H1 (stars) of 
the scaled longitudinal momentum $x_F$ (left) and of 
the average transverse momentum squared (with respect to the 
photon-pomeron axis) as a function of $x_F$, for charged particles (right).
The curves are model predictions.}
\label{fig:ZEUS_part}
\end{figure}

Both HERA experiments have also measured the momentum distributions of
charged particles in the forward (photon) and backward (pomeron) hemispheres
of diffractive events
% (Fig. \ref{fig:ZEUS_part}a)
\cite{ZEUS_final_state,H1_final_state}. 
These spectra have been compared by H1 to those obtained in fixed target 
non-diffractive DIS lepton-proton interactions in the same kinematic range
($W^{DIS} \simeq M_X^{LRG}$).
Whereas in DIS a strong asymmetry is observed between the photon hemisphere 
and the proton remnant region, characterised by a reduced particle emission, 
the momentum spectra are similar for both hemispheres of LRG events
(see Fig. \ref {fig:ZEUS_part}, left \cite{H1_final_state}), and they are softer 
than for DIS.
%, which is attributed to the increased central particle emission 
%in the octet-octet case of $q \bar q g} events
% $-$ or similarly to the role of the 
%BGF process in the case of a gluon-dominated pomeron, as confirmed by the 
%analysis of RAPGAP predictions: a good description is obtained when the 
%parton distributions obtained from the inclusive cross section are used, but 
%not for a $q \bar q$ pomeron \cite{H1_final_state}.
%The ``sea-gull'' plots, 
%(Fig. \ref{fig:ZEUS_part}b) leads to similar conclusions.
The distribution of the average transverse momentum squared (with respect 
to the photon direction in the $X$ system rest frame) as a function 
of $x_F$ (``sea-gull'' plots $-$ Fig. \ref {fig:ZEUS_part}, right),
show that charged particles in LRG events have larger transverse
momentum than in DIS events.
These features point towards a significant role of photon fluctuation 
topologies of the type $q \bar q g$ or, equivalently, towards a gluonic pomeron.

\begin{figure}[tbp]
\vspace{0.cm}
\begin{center}
  \setlength{\unitlength}{1.0cm}
  \begin{picture}(16.0,8.0)
  \put(0.0,0.0){\epsfig{file=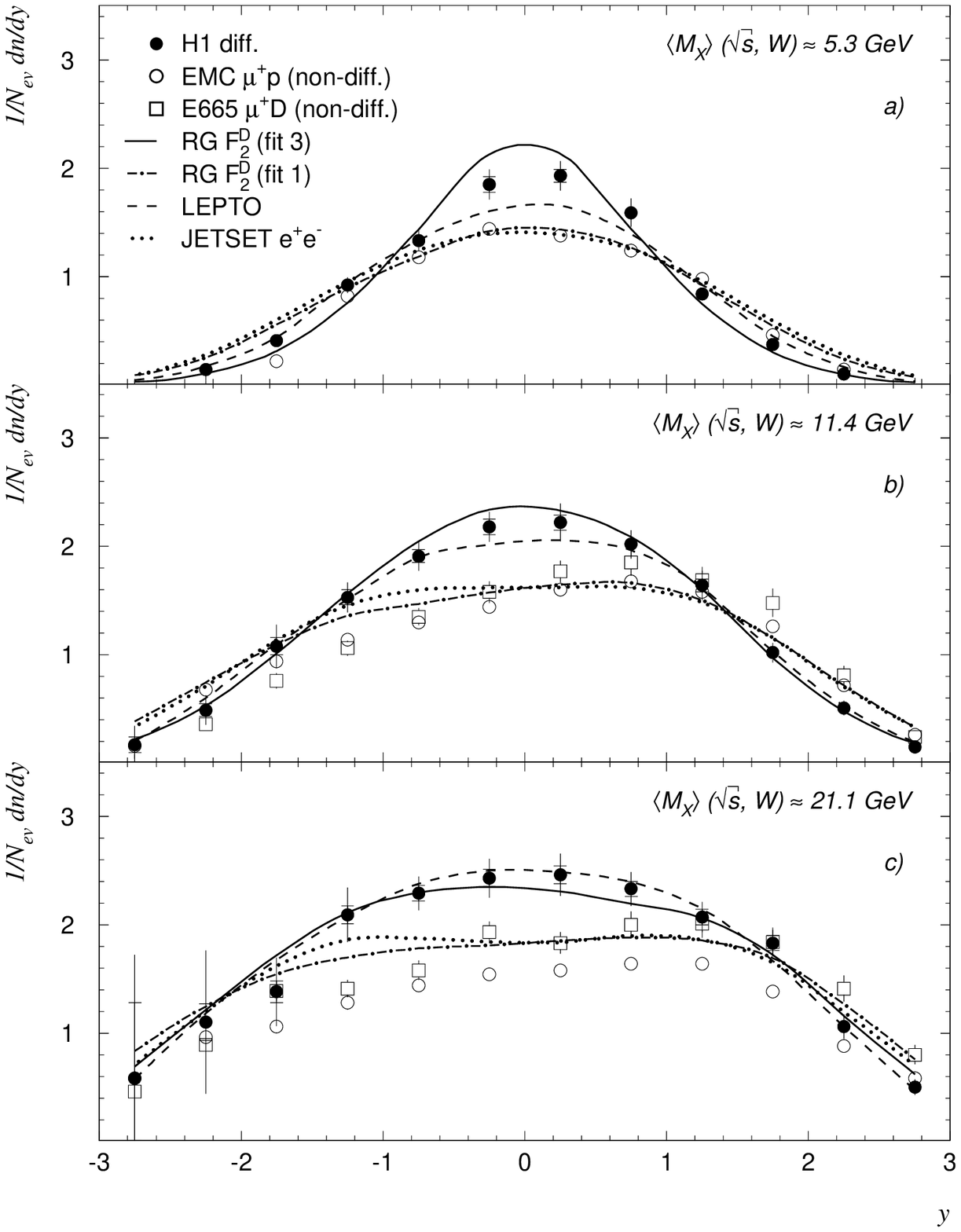,width=8cm,height=8cm}}
  \put(8.0,0.0){\epsfig{file=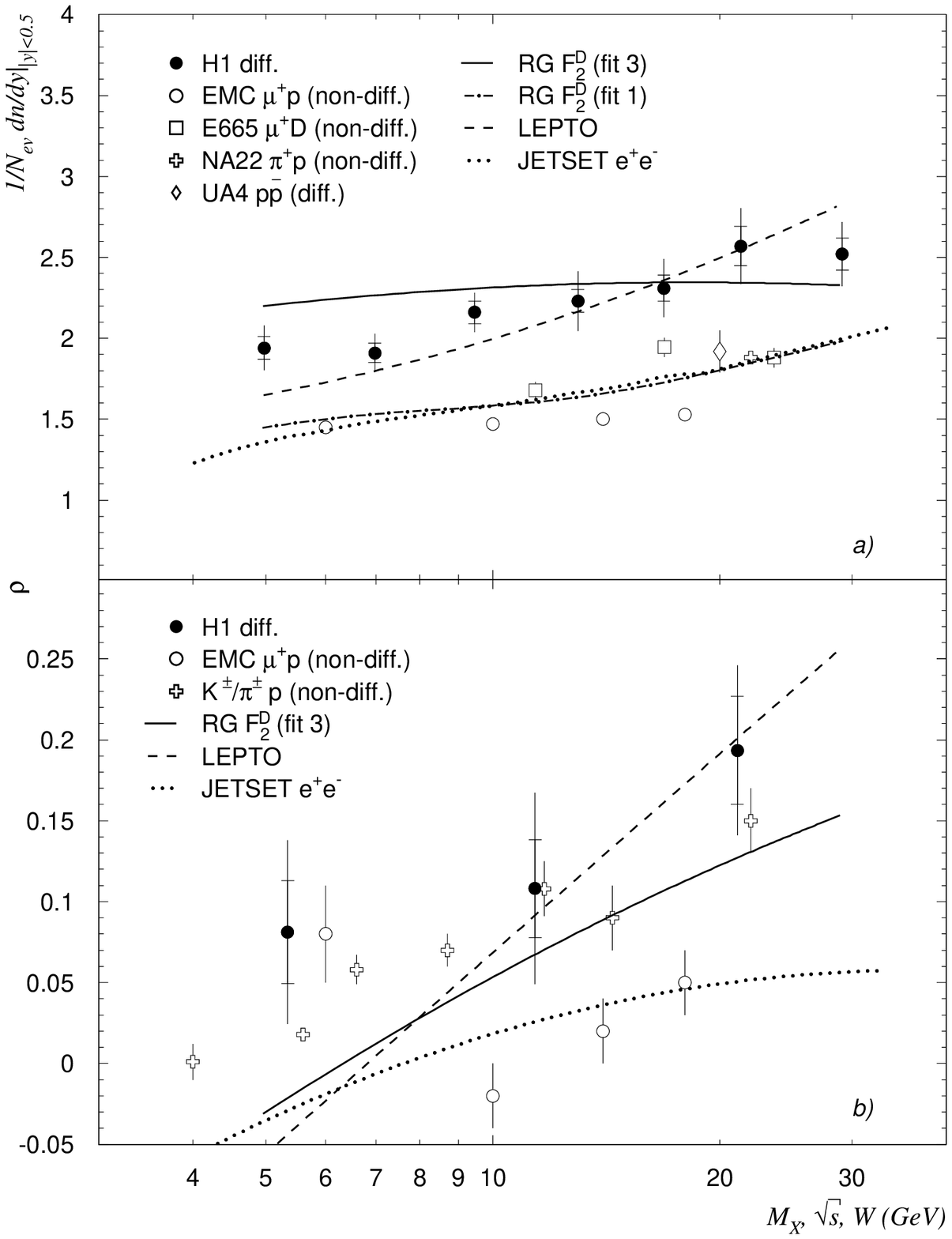,width=8cm,height=8cm}}
  \put(8.0,4.3){\epsfig{file=whitebox.eps,width=8cm,height=4cm}}
  \end{picture}
\end{center}
\vspace{-0.cm}
\caption{Left: charged particle multiplicity distribution of DIS LRG events,
measured by H1 for three intervals of $M_X$, compared to non-diffractive 
lepton$-$proton 
data and to $e^+e^-$ interactions, represented by the JETSET calculation. 
Right: correlations between the backward and forward hemispheres for H1 LRG events,
compared to non-diffractive lepton$-$proton and meson$-$proton data and to 
$e^+e^-$ interactions, represented by the JETSET calculation.
The curves are predictions of several Monte Carlo models:
RAPGAP with the parton distributions extracted from a DGLAP fit to the inclusive
diffractive cross section (``fit 3'') or with quarks only at the starting scale 
(``fit 1''), LEPTO, JETSET.}
\label{fig:H1_eflow_mult}
\end{figure}

Multiplicity \cite{H1_mult} and energy flow \cite{H1_final_state} studies 
reinforce these conclusions.
The central charged particle multiplicity for diffractive events is 
significantly higher than for $e^+ e^-$ events 
(see Fig. \ref{fig:H1_eflow_mult}, left), which is attributed to 
stronger parton radiation in the effective octet-octet structure of diffraction 
than for the triplet-triplet interactions of $e^+ e^-$ data.
The role of gluons is confirmed by the comparison of the data with predictions 
of the RAPGAP model:
the central activity, both for particle multiplicity and energy flow, is well 
described by the RAPGAP Monte Carlo when using 
the parton distributions as obtained from the DGLAP fit to the inclusive cross 
section measurement, whereas the prediction is significantly 
too low for a hypothetic $q \bar q$ pomeron.
Finally, the correlations in the multiplicity distributions between the backward 
and forward hemispheres in diffractive DIS (Fig. \ref{fig:H1_eflow_mult}, right)
are stronger than for $e^+ e^-$ or non-diffractive lepton hadron interactions, 
which correspond to triplet-triplet topologies.
In contrast, they are of comparable strength to those in soft hadron 
interactions, where the strong correlations are attributed to the number
of overlapping strings in phase space.

It must be noted that most features of the diffractive final states are 
not only well reproduced by a gluon dominated pomeron as implemented in RAPGAP, 
but also by the LEPTO SCI model.
This is consistent with the major role attributed to gluons in diffraction, 
since the low $x$ parton distribution functions, with an important gluon 
contribution, are input to the LEPTO model.

%=======================\label{fig:ZEUS_part} =================
%=======================\end{fig:ZEUS_part} =================

%=======================\label{fig:H1_eflow_mult} =================
%=======================\end{fig:H1_eflow_mult} =================

%=========================================================
%=========================================================
\subsection {Jet and Charm Production}
						
							\label{sect:jets}

The studies of inclusive properties of the diffractive final state are
complemented by semi-inclusive studies of the characteristics of jet and
charm production in diffraction.

The particular interest of these processes is in the presence of a well defined 
hard scale (jet $p_t$ or charm mass), which provides an access to perturbative
QCD calculations.
Unfortunately, the cross section for jet production is small, and 
stringent experimental selection procedures have to be imposed to select
open charm events.
The full potential of these processes for understanding hard diffraction has 
thus not been exploited yet.

Jets with transverse momenta larger than 5 (H1) or 6 GeV (ZEUS), defined
with respect to the photon direction in the rest frame of the system $X$, have
been studied in photoproduction both by H1 \cite{H1_jets} and 
ZEUS \cite{ZEUS_jets}, and in DIS by H1 \cite{H1_jets}.

\begin{figure}[tbp]
\begin{center}
\vspace{-1.5cm}
 \setlength{\unitlength}{1.0cm}
 \begin{picture}(18.0,10.0)
  \put(0.0,-1.0){\epsfig{file=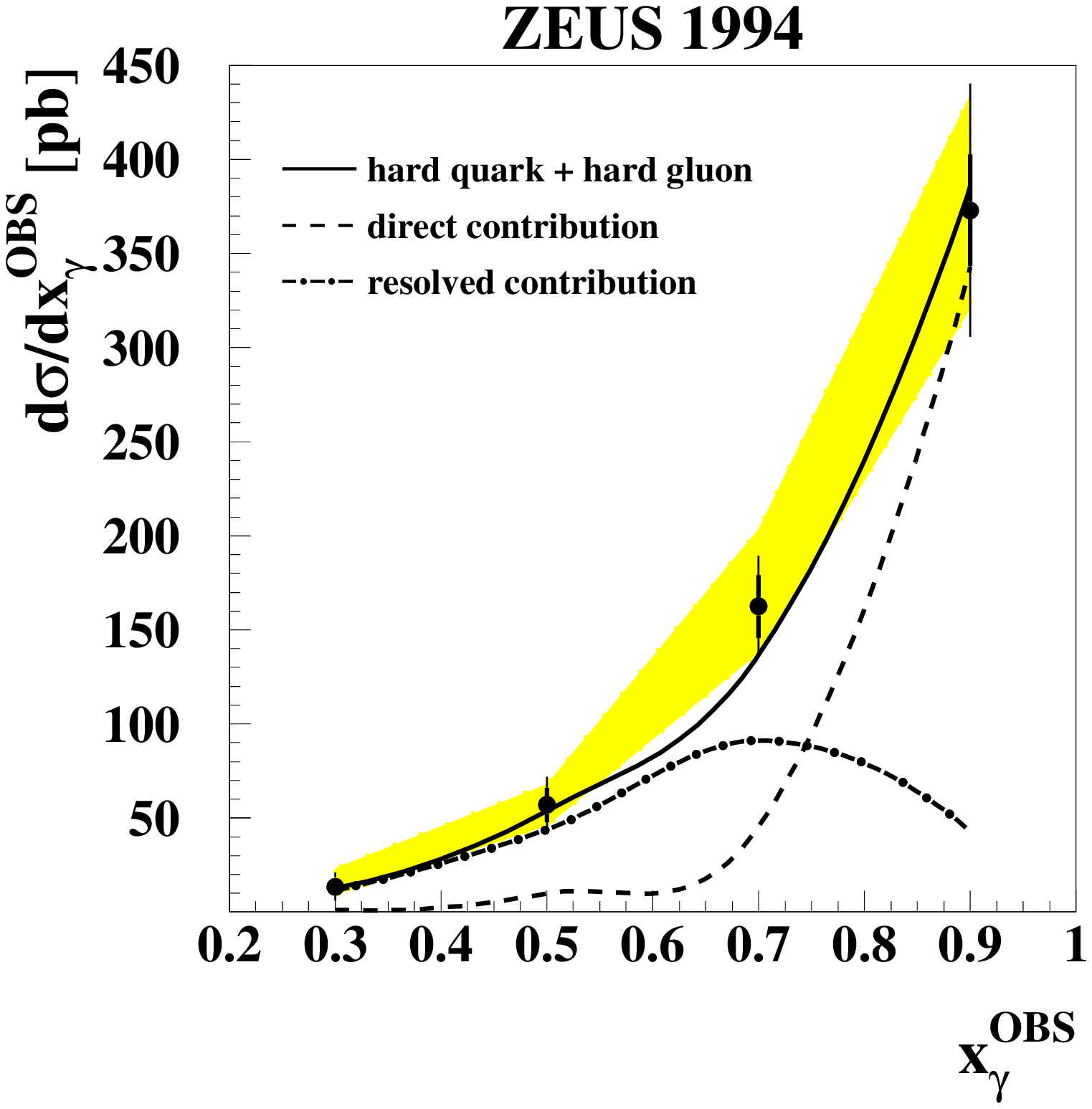,width=9cm,height=10cm}}
  \put(9.0,4.5){\epsfig{file=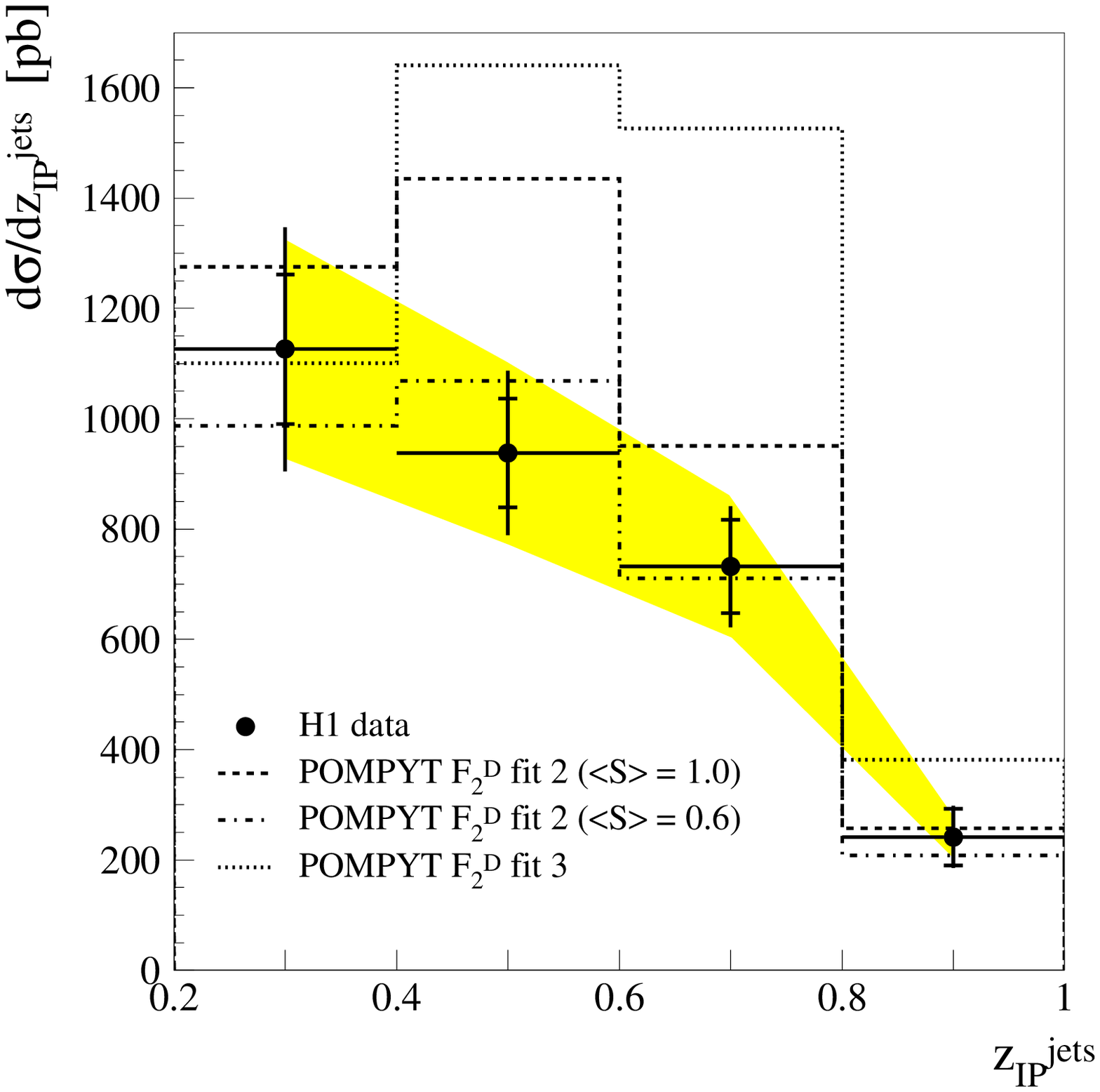,width=9cm,height=5cm}}
  \put(9.0,0.0){\epsfig{file=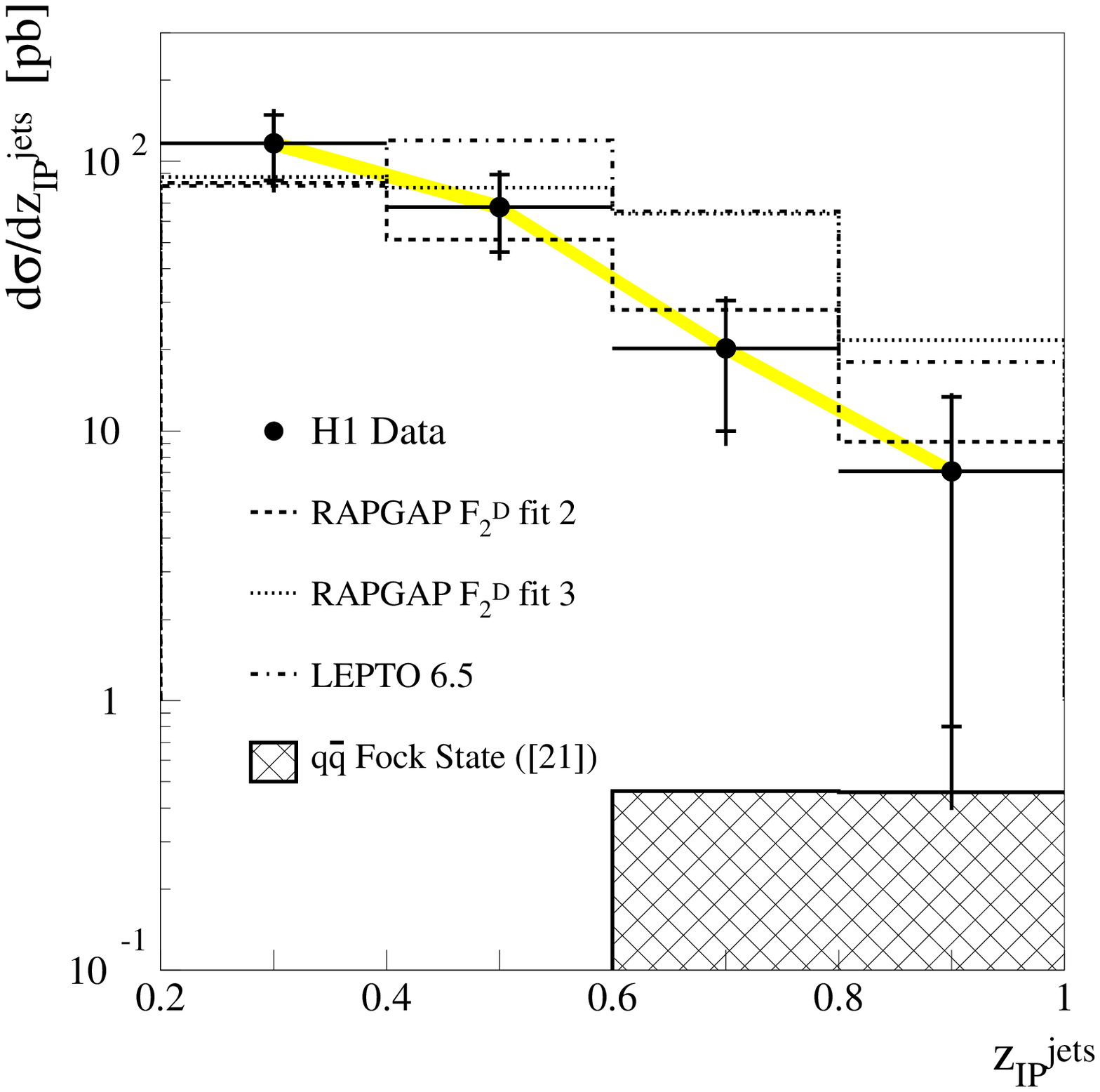,width=9cm,height=5cm}}
 \end{picture}
\end{center}
\vspace{0.cm}
\caption{
Left: ZEUS measurement of the cross section of diffractive dijet 
photoproduction for 
$E_T^{jet} > 6$ \gev\ and $-1.5 < \eta^{jet} < 1$, as a function of $x_\gamma$, 
the fraction of the photon momentum entering the hard interaction; 
the shaded area represents a systematic uncertainty due to the absolute energy 
scale of the jets; 
the curves represent the resolved (dot-dashed line), the direct
(dashed line) and the resolved + direct contributions, based on the pomeron
parton distribution parameterised as hard quarks and hard gluons, in a joint 
QCD fit of the inclusive DIS diffractive cross section and the dijet diffractive 
photoproduction.
Right: H1 measurements of the cross section of diffractive dijet 
photoproduction (top) and of DIS production (bottom) for
$p_T^{jet} > 5$ \gev\ and $-1 < \eta^{jet} < 2$, as a function of $z$, 
the fraction of the pomeron momentum entering the hard interaction;
the shaded bands show the overall normalisation uncertainties;
the data are compared to the predictions of the POMPYT (photoproduction) and
RAPGAP (DIS) Monte Carlo models with parton densities dominated by a ``flat''
(``fit 2'') or ``peaked'' (``fit 3'') gluon distribution at the starting 
scale; in the photoproduction case, the POMPYT prediction for the ``flat''
distribution is also shown for a survival probability of 0.6; in the DIS case,
the prediction of a $q \bar q$ Fock state calculation is shown as well.}
\label{fig:jets}
\end{figure}

In photoproduction, the distribution of the $x_\gamma$ variable, which 
describes the fraction of the photon momentum entering the hard interaction,
requires the presence of both a direct ($x_\gamma \simeq 1$) and a resolved 
($x_\gamma < 1$) contribution (see Fig. \ref{fig:jets}, left).

Both for photoproduction and DIS jet production, a sizeable contribution is
observed of events where a large fraction $z$ of the pomeron momentum enters 
the hard interaction; however, the cross section increases as $z$
decreases, suggesting the presence of pomeron remnants 
(see Fig. \ref{fig:jets}b, right).
This supports the hypothesis of a small jet cross section for a $q \bar q$ 
Fock state, due to screening effects (see section \ref{sect:qual_features}), 
as demonstrated by the fact that the predictions of a relevant model 
\cite{Bartels_jets} and of the RAPGAP Monte Carlo with a $q \bar q$ pomeron 
at the starting scale are significantly too small.
The bulk of the data is thus explained by $q \bar q g$ states, as expected 
both in perturbative QCD calculations (see e.g. \cite{NNN92,evol}) and in a 
semi-classical approach of soft colour interactions \cite{semicl_jets}. 

The results of both experiments support models where the partonic structure
of the pomeron is dominated by hard gluons.
This is shown by ZEUS using a simultaneous fit of the inclusive diffractive 
DIS cross section and of the jet production cross section, and by H1
by implementing in the RAPGAP calculation the parton densities obtained
from the fits to the scaling violations in inclusive DIS.
H1 observes that parton distributions in which the pomeron gluon structure
is relatively flat (``fit 2'' in \cite{H1_f2d_94}) describe the data better
than those in which the gluon distribution is peaked at large $z$ (``fit 3'').
In resolved photoproduction, a possible underlying interaction between the
proton and the photon remnant can be parameterised as a ``survival
probability'' \cite{survival_proba}.
The best description of the combined DIS and photoproduction data is obtained
when a rapidity gap survival probability of 0.6 
is applied to the ``flat'' gluon distribution, but the measurements are
affected by large uncertainties.

%=======================\label{fig:jets} =================
%=======================\end{fig:jets} =================

%=======================\label{fig:rapgap_jets} =================
\begin{figure}[tbp]
\begin{center}
\vspace{0.cm}
 \setlength{\unitlength}{1.0cm}
 \begin{picture}(12.0,6.0)
  \put(-6.0,0.0){\epsfig{file=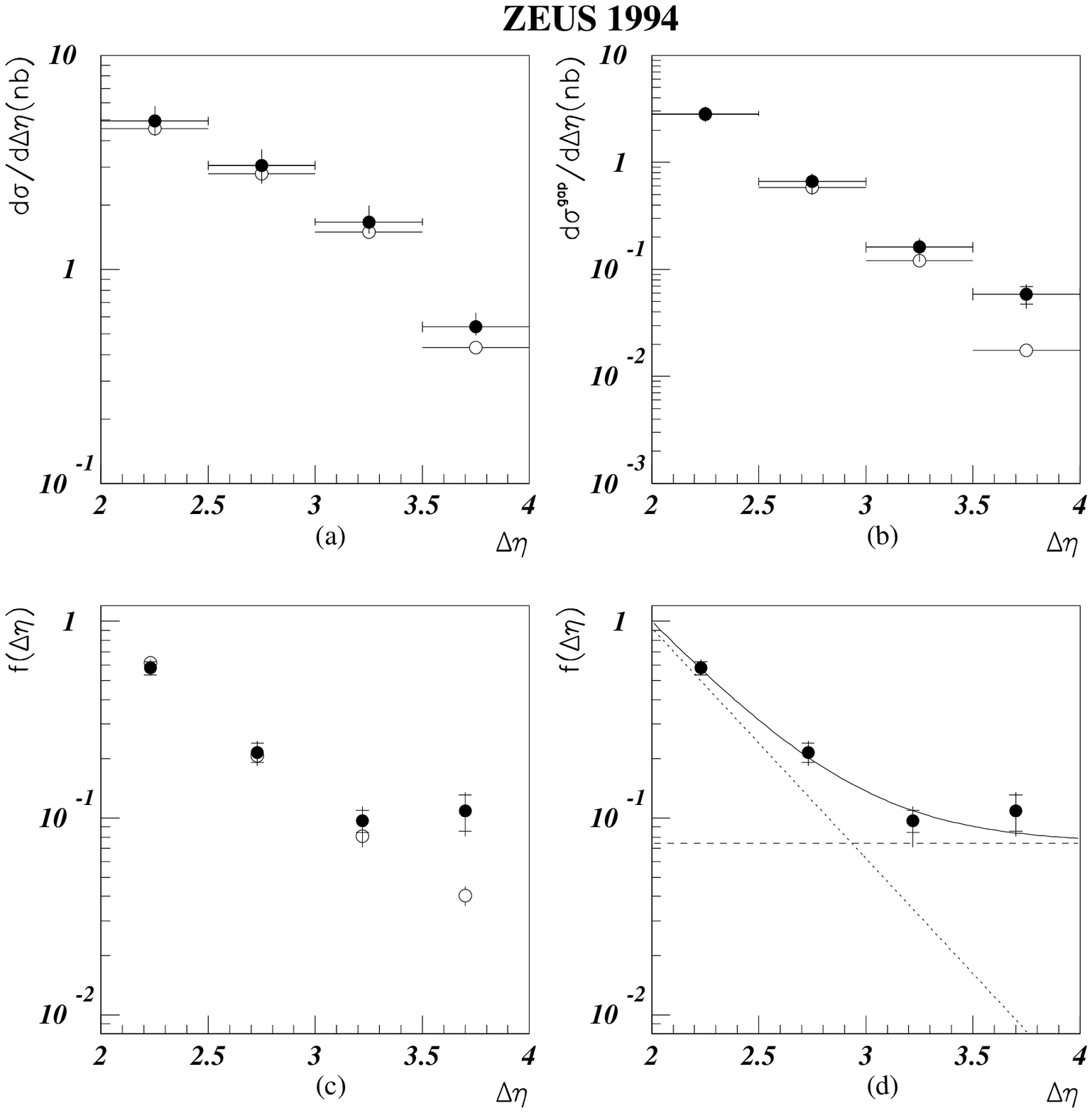,width=12cm,height=12cm}}
  \put(6.0,-0.4){\epsfig{file=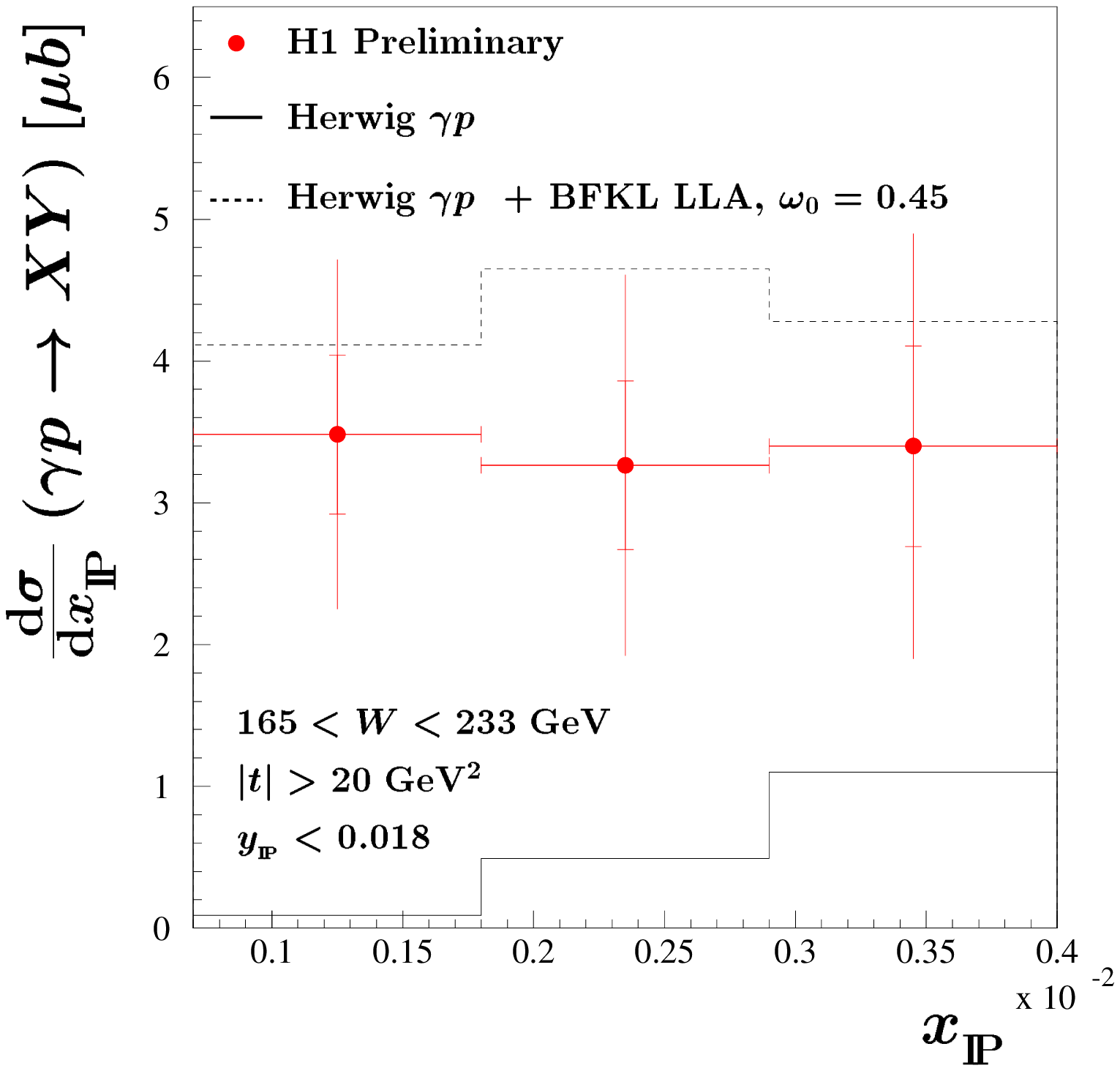,width=6.6cm,height=6.6cm}}
  \put(-6.0,-1.0){\epsfig{file=whitebox.eps,width=6cm,height=7cm}}
  \put(-6.0,6.){\epsfig{file=whitebox.eps,width=13cm,height=6cm}}
 \end{picture}
\end{center}
\vspace{0.5cm}
\caption{Left: ZEUS measurement of the fraction of dijet photoproduction 
events separated by the interval $\Delta \eta$ in rapidity, with no hadronic 
activity between the jets, as a function of $\Delta \eta$;
the distribution is fitted as the sum of an exponentially falling 
contribution (dotted line), attributed to events with ``usual'' colour
exchange properties, and of a constant contribution (dashed line), attributed
to colour-singlet exchange between the jets.
Right: H1 measurement of the cross section, differential in \xpom, of
photoproduction events with four-momentum transfer squared $|t| > 20$ \gevsq;
the solid line is the prediction of a standard photoproduction Monte Carlo;
the dashed line is the prediction of a model based on perturbative QCD
calculations (with well defined slope but uncertain absolute normalisation).}
\label{fig:rapgap_jets}
\end{figure}
%=======================\end{fig:rapgap_jets} =================

Diffractive exchange is also manifest in a class of
events containing jets, and characterised by the absence of hadronic activity
between the jets.
In photoproduction, both ZEUS and H1 have observed a signal for such
events, in excess over the exponential fall-off expected from ``usual''
colour exchange and hadronisation properties
(see Fig. \ref{fig:rapgap_jets}, left).
This signal is  attributed to the exchange between the hard partons producing
the jets of a coulour-singlet object, presumably the pomeron.

Following a suggestion to relax the requirement of observing the two jets in 
the main detector \cite{cox_forshaw}, the H1 collaboration has selected 
photoproduction events characterised by the presence of a gap in rapidity
of at least 1.5 units between two systems, $X$ and $Y$, of masses 
$M_X, M_Y \ll W$, with $|t| = p_{tX}^2 > 20$ \gevsq \cite{H1_high_t}.
In the selected kinematics range, a significant excess of events is observed
above the expectation for standard photoproduction processes
(see Fig. \ref{fig:rapgap_jets}, right).
The measurement is performed differentially in \xpom, and allows direct
comparison with predictions based on perturbative QCD calculations 
\cite{cox_forshaw}: although the absolute nor;alisation is uncertain, the
slope is well described, in contrast with the case of ``standard'' Monte Carlo
simulations.

%=======================\label{fig:charm} =================
\begin{figure}[tbp]
\vspace*{0.0cm}
\begin{center}
\epsfig{file=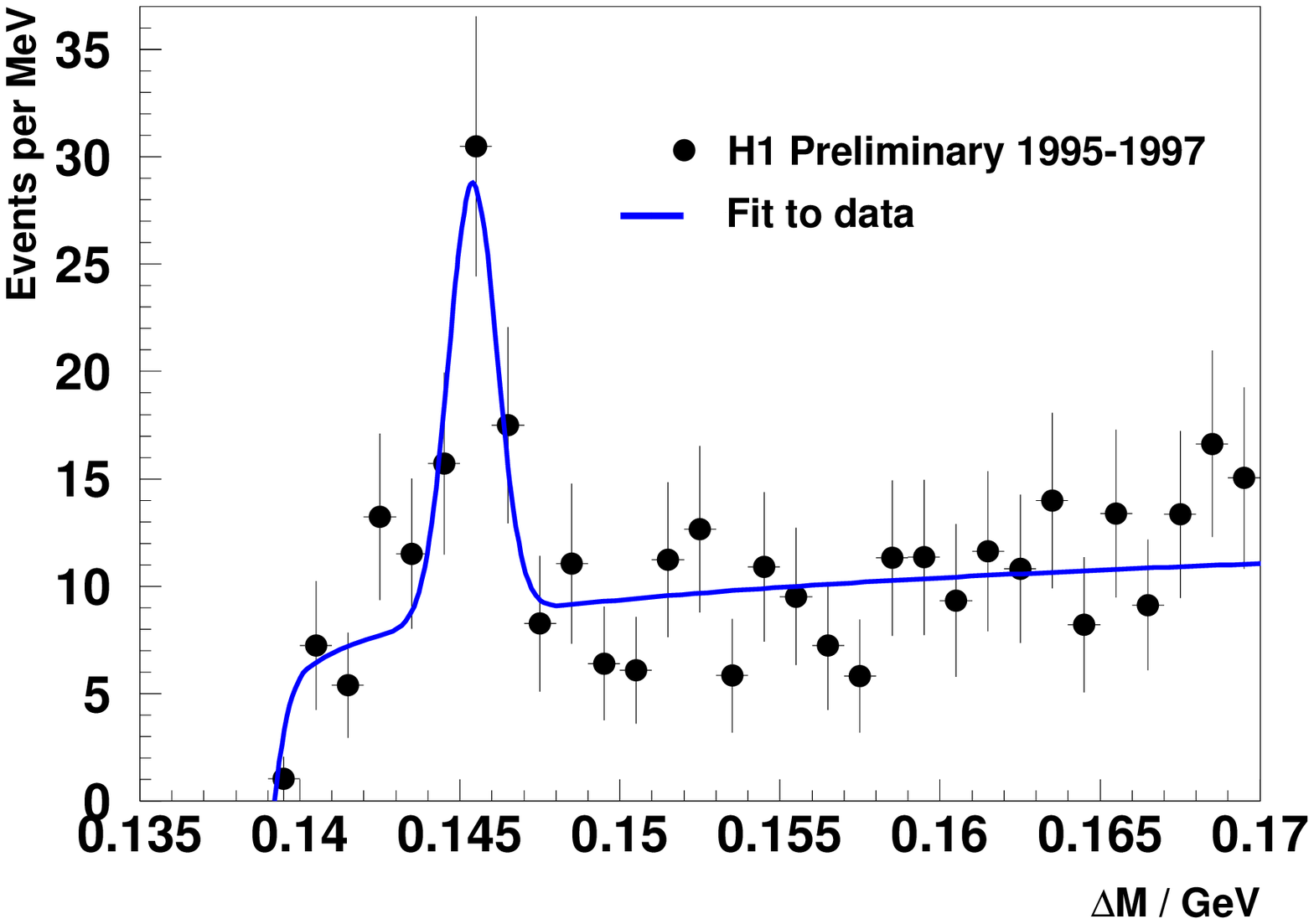,width=8cm,height=4cm}
\epsfig{file=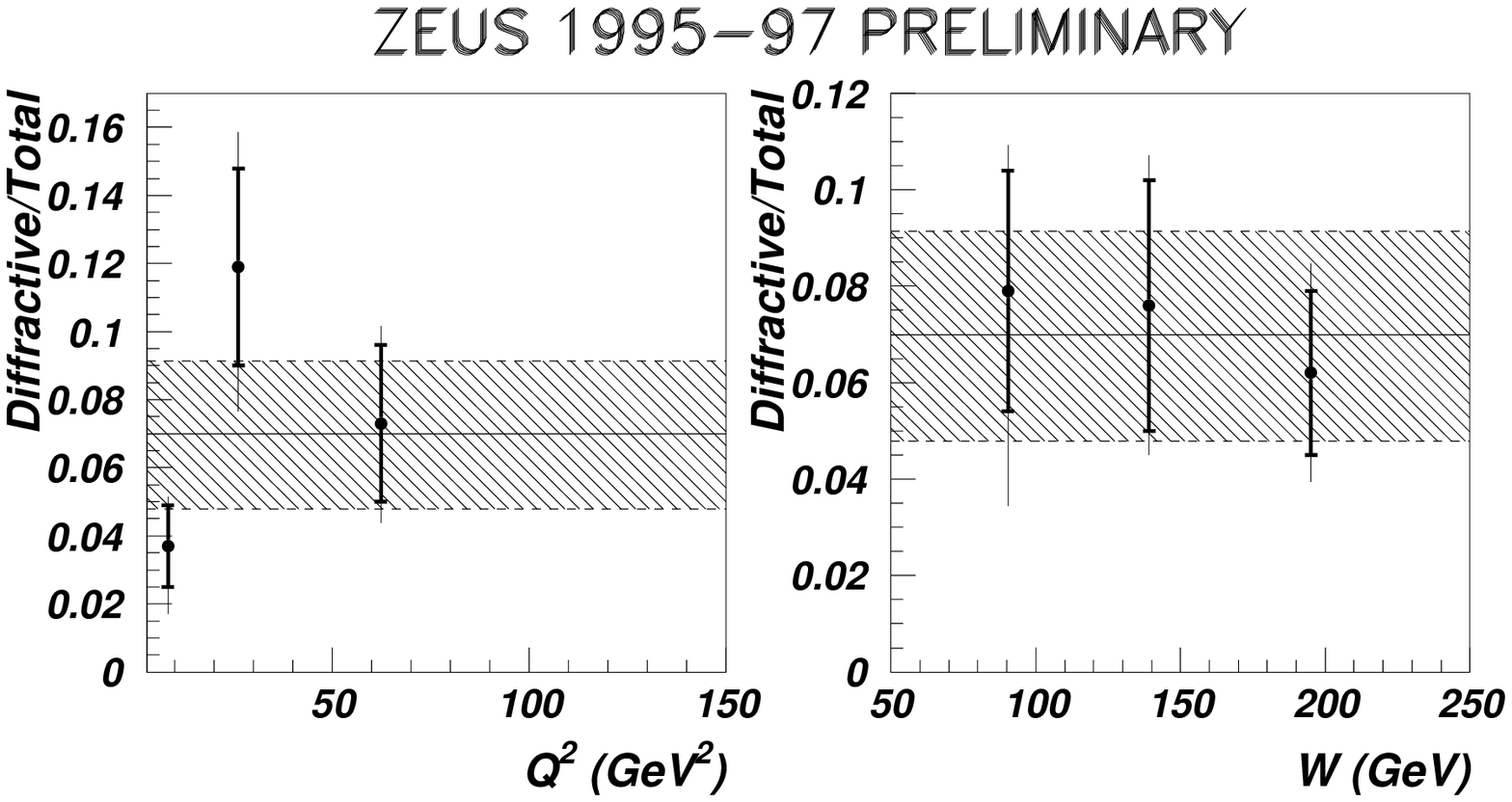,width=8cm,height=5cm}
\end{center}
\vspace{.cm}
\caption{Left: H1 measurement of the mass difference 
$\Delta M = M(K^- \pi^+ \pi^+_s) - M(K^- \pi^+)$, exhibiting a clear signal
for diffractive DIS charm production, of $38 \pm 10 \pm 4$ events.
Right: ZEUS measurement of the fraction of diffractively produced $D^{*\pm}$ 
mesons, as a function of \qsq\ and $W$.}
\label{fig:charm}
\end{figure}
%=======================\end{fig:charm} =================

Both experiments have also reported the observation of a signal for 
diffractive production of open charm in DIS, in the channel 
$D^{*+} \rightarrow D^0 \pi^+_s \rightarrow (K^- \pi^+) \pi^+_s$ (and the charge
conjugate), where the particle noted $\pi^+_s$ is a slow pion
\cite{H1_charm,ZEUS_charm} (see Fig. \ref{fig:charm}, left).
The fraction of diffractively produced $D^{*\pm}$ mesons is measured by
ZEUS to be of $7.0 \pm 1.3 \pm 1.7 \%$, which is consistent with the
fraction of the total DIS cross section attributed to diffraction
(see Fig. \ref{fig:charm}, right).
The study of charm production is a promising field to discriminate between 
several models of diffraction \cite{charm}.
An important role is assumed to be played by the gluonic content of the
pomeron, charm being dominantly produced through the boson-gluon fusion process.

%=========================================================
%=========================================================
\section {Exclusive Vector Meson Production}
							\label{sect:VM}

Abundant and very interesting data have been accumulated on exclusive
vector meson production, which have only partly been interpreted theoretically.

The most striking feature in this field is the observation of a fast increase 
with energy of the photoproduction cross section of \jpsi\ mesons.
This increase can be related to the fast increase at low $x$ of the gluon 
density in the proton, and the understanding of these data is probably the 
greatest theoretical success so far of perturbative QCD studies in the whole 
field of diffraction.
It is true to say that the low $x$ behaviour of the proton structure function 
and the energy dependence of \jpsi\ photoproduction are the two definite 
evidences of a ``hard'' behaviour of strong interactions at HERA.

As this rich sample of experimental data has been reviewed in detail in the 
present Conference \cite {Whitmore,Crittenden,West}, it will not be discussed 
in this introduction.

%=========================================================
%=========================================================
\section {Conclusions}
      							\label{sect:concl}

In the last few years, high energy diffraction has been the subject of great 
interest, as testified by a large number of publications, both experimental 
and theoretical, and of specialised workshops, where experimentalists and 
theorists actively interact.

The large amount of results accumulated at HERA, complemented by data 
collected at the Tevatron,  have renewed the experimental approaches:
total cross section measurements have been complemented by studies of the 
inclusive diffractive final states in the deep-inelastic regime 
(charged particle distributions
and multiplicities, energy flow, event shape), of diffractive jet and charm
production, and of exclusive vector meson production.
On the theoretical side, in the light of the experimental results,
emphasis is placed on the interpretation of diffraction in terms of QCD.

Inclusive interactions of real photons with protons and
light vector meson photoproduction are governed, as for hadron--hadron total 
cross section, by ``soft'' diffraction, with a mild energy dependence of the
cross section.
On the other hand, high \qsq\ deep-inelastic scattering and \jpsi\ exclusive 
production (both by real and virtual photons) reveal a strong energy 
dependence of the cross section, characteristic of ``hard'' processes, 
related to a fast increase of the gluon content in the proton at high
energy (low $x$ values).
As for  total diffractive cross section and light vector meson exclusive
electroproduction (with \qsq\ $\gsim $ a few \gevsq ), they present evidence 
for an interplay between hard and soft diffraction, i.e. between perturbative 
and non-perturbative QCD features.

The available data are consistent with the interpretation of diffraction
as due to the exchange in the $t$-channel of a gluon dominated object.
A DGLAP analysis of scaling violations of the total diffractive cross section,
in the form of the \fdthree\ structure function (possibly complemented by
a large higher twist contribution at large $\beta$), favours the dominance of 
hard gluons in the pomeron.
Assuming factorisation of the cross section into a pomeron flux in the proton
and a hard scattering process, the corresponding parton distributions, 
propagated e.g. through the RAPGAP calculation, give
a good description of the inclusive final states and of jet production, 
whereas the data are inconsistent with a quark dominated pomeron.
Seen from the proton rest frame, the data can be interpreted as due mainly to 
the fluctuation of the photon in a $q \bar q g$ Fock state, a 
$q \bar q$ system leading to mutual colour screening of the quarks at
large $p_T$ (i.e. for small distances).

It should be noted that many features of the experimental data can also be
reproduced by models (specifically the LEPTO calculation) which do not
refer to diffraction as a specific process but use the concept of soft colour
interactions in the proton, the hard interaction being driven by the gluon 
dominance in the proton at small $x$.

More theoretical work is needed to progress in the understanding of diffraction
in terms of QCD.
In particular, it is important to formulate predictions 
for the most accessible experimental characteristics of diffractive events:
event shape, energy flow, multiplicity distributions, as well as for jet and
charm production.

On the experimental side, a better understanding of the data is necessary in
view of resolving the remaining discrepancies between H1 and ZEUS for the 
total cross section measurement.
Major progress is expected from a significant increase of statistics, 
in particular for the study of 
hard diffraction: jet and charm production, high \qsq\ ($ \gsim 20 $ \gevsq )
vector meson production, and particularly high $t$ interactions, for which the 
present data are very limited.
In addition, a measurement of the diffractive longitudinal cross section would
be a precious tool to discriminate between models and to
specify the domain of applicability of DGLAP evolution equations.

A bright future is open for the study of diffraction and its understanding in
terms of QCD.

%
%
%
%====================================================
%====================================================
%

\end{document}